\documentclass[ twocolumn, tigthen, times]{aastex631}

\shorttitle{JWST MIRI  Caldetector 1 steps}
\shortauthors{Morrison et al.}

\usepackage{hyperref}
\begin{document}

\title{JWST MIRI flight performance: Detector Effects and Data Reduction Algorithms}

\author[0000-0002-9288-9235]{Jane E.\ Morrison}
\affiliation{Steward Observatory, University of Arizona, Tucson, AZ, 85721, USA}

\author[0000-0003-0589-5969]{Daniel Dicken}
\affiliation{UK Astronomy Technology Centre, Royal Observatory Edinburgh, Blackford Hill, Edinburgh EH9 3HJ, UK}

\author[0000-0003-2820-1077]{Ioannis Argyriou}
\affiliation{Institute of Astronomy, KU Leuven, Celestijnenlaan 200D, 3001 Leuven, Belgium}

\author[0000-0001-5644-8830]{Michael E.\ Ressler}
\affiliation{Jet Propulsion Laboratory, California Institute of Technology, 4800 Oak Grove Dr., Pasadena, CA, 91109, USA}

\author[0000-0001-5340-6774]{Karl D.\ Gordon}
\affiliation{Space Telescope Science Institute, 3700 San Martin Drive, Baltimore, MD, 21218, USA}

\author[0000-0001-9367-0705]{Michael W.\ Regan}
\affiliation{Space Telescope Science Institute, 3700 San Martin Drive, Baltimore, MD, 21218, USA}

\author[0000-0002-7698-3002]{Misty Cracraft}
\affiliation{Space Telescope Science Institute, 3700 San Martin Drive, Baltimore, MD, 21218, USA}

\author[0000-0003-2303-6519]{George H.\ Rieke}
\affiliation{Steward Observatory, University of Arizona, Tucson, AZ, 85721, USA}

\author[0000-0003-0209-674X]{Michael Engesser}
\affiliation{Space Telescope Science Institute, 3700 San Martin Drive, Baltimore, MD, 21218, USA}

\author[0000-0002-8909-8782]{Stacey Alberts}
\affiliation{Steward Observatory, University of Arizona, Tucson, AZ, 85721, USA}

\author[0000-0002-7093-1877]{Javier Alvarez-Marquez} \affiliation{Centro de Astrobiología (CAB, CSIC-INTA), Carretera de Ajalvir, 28850 Torrejón de Ardoz, Madrid, Spain}

\author[0000-0001-6482-3020]{James W.\ Colbert}
\affiliation{IPAC, Mail Code 314-6, Caltech, 1200 E. California Blvd., Pasadena, CA, 91125, USA}

\author[0000-0003-2238-1572]{Ori D.\ Fox}
\affiliation{Space Telescope Science Institute, 3700 San Martin Drive, Baltimore, MD, 21218, USA}
  
\author[0000-0002-1257-7742]{Danny Gasman}
\affiliation{Institute of Astronomy, KU Leuven, Celestijnenlaan 200D, 3001 Leuven, Belgium}

\author[0000-0002-9402-186X]{David R.\ Law}
\affiliation{Space Telescope Science Institute, 3700 San Martin Drive, Baltimore, MD, 21218, USA}

\author[0000-0003-4801-0489]{Macarena Garcia Marin}
\affiliation{Space Telescope Science Institute, 3700 San Martin Drive, Baltimore, MD, 21218, USA}

\author[0000-0001-8612-3236]{Andr\'{a}s G\'{a}sp\'{a}r}
\affiliation{Steward Observatory, University of Arizona, Tucson, AZ, 85721, USA}

\author[0000-0002-2421-1350]{Pierre Guillard} 
\affiliation{Sorbonne Universit\'{e}, CNRS, UMR 7095, Institut d'Astrophysique de Paris, 98bis bd Arago, 75014 Paris, France}
\affiliation{Institut Universitaire de France, Minist{\`e}re de l'Enseignement Sup{\'e}rieur et de la Recherche, 1 rue Descartes, 75231 Paris Cedex 05, France}

\author[0000-0002-7612-0469]{Sarah Kendrew}
\affiliation{European Space Agency, at Space Telescope Science Institute, 3700 San Martin Drive, Baltimore, MD, 21218, USA}

\author[0000-0002-0690-8824]{Sarah Kendrew}
\affiliation{Telespazio UK for the European Space Agency, ESAC, Camino Bajo del Castillo s/n, 28692 Villanueva de la Ca\~nada, Spain.}
\affiliation{ETH Zurich, Institute for Astronomy, Wolfgang-Pauli-Str. 27, CH-8093 Zurich, Switzerland}

\author[0000-0003-1250-8314]{Seppo Laine}
\affiliation{IPAC, Mail Code 314-6, Caltech, 1200 E. California Blvd., Pasadena, CA, 91125, USA}

\author[0000-0002-6296-8960]{A.\ Noriega-Crespo}
\affiliation{Space Telescope Science Institute, 3700 San Martin Drive, Baltimore, MD, 21218, USA}

\author[0000-0003-4702-7561]{Irene Shivaei}
\affiliation{Steward Observatory, University of Arizona, Tucson, AZ, 85721, USA}

\author[0000-0003-4520-1044]{G.~C.\ Sloan}
\affiliation{Space Telescope Science Institute, 3700 San Martin Drive, Baltimore, MD, 21218, USA \& \\
Department of Physics and Astronomy, University of North Carolina, Chapel Hill, NC 27599-3255, USA}

\begin{abstract}
The detectors in the Mid-Infrared Instrument (MIRI) of the James Webb Space Telescope (JWST) are arsenic-doped silicon impurity band conduction (Si:As IBC) devices and are direct descendants of the Spitzer IRAC long wavelength  arrays (channels 3 and 4). With appropriate data processing, they can provide excellent performance.   In this paper we discuss the various non-ideal behaviors of these detectors that need to be addressed to realize their potential.  We have developed a set of algorithms toward this goal, building on experience with previous similar detector arrays.   The MIRI-specific stage 1 pipeline algorithms, of a three stage JWST calibration pipeline, were developed using pre-flight tests on the flight detectors and flight spares and have been refined using flight data.  This paper describes these algorithms, which are included in the first stage of the  JWST Calibration Pipeline for the MIRI instrument.  
\end{abstract}


\section{Introduction} \label{sec:intro}
 The Mid-Infrared Instrument (MIRI) for JWST \citep{wright2023} uses three arsenic-doped silicon impurity band conduction (Si:As IBC) detectors, one in the imaging module 
(\cite{Dicken2023}; \cite{Kendrew2023}), and two in the medium-resolution spectrometer (MRS) (\cite{Argyriou2023}).
These types of detectors have a long and successful history in space missions. Si:As IBC arrays were used on all three {Spitzer} instruments (Infrared Array Camera: \citeauthor{fazio2004} \citeyear{fazio2004}; \citeauthor{hora2004} \citeyear{hora2004}; Infrared Spectrograph: \citeauthor{vancleve1995} \citeyear{vancleve1995}; \citeauthor{houck2004} \citeyear{houck2004}; and Multi-Band Imaging Photometer for Spitzer: \citeauthor{rieke2004}  \citeyear{rieke2004}; \citeauthor{gordon2005} \citeyear{gordon2005}). They have also been used successfully in many other infrared space missions, for example the Wide-field Infrared Survey Explorer  \citep[WISE,][]{mainzer2008}, the Midcourse Space Experiment \citep[MSX,][]{mill1994}, and Akari \citep{onaka2007}.  For all three Spitzer instruments, the Si:As IBC detectors proved very stable and capable of reproducible measurements to the 1\% level. 

A number of papers have been written detailing the design and operation of the MIRI detectors. For the purposes of this paper we will highlight the aspects of detector array, readout electronics, and focal plane system that affect the MIRI algorithms in stage 1 of the JWST calibration pipeline, the {\sc calwebb\_detector1} pipeline. The full three stages of the pipeline are discussed by \cite{jwst_pipeline2022}. For details on the design and operation of these types of detectors see \citet{rieke2015}. For information on the readout integrated circuit (ROIC), electronic data chain, mechanical arrangement, and operation of the MIRI arrays, see \citet{ressler2015}.

\subsection{Detectors and Focal Plane System}\label{FPS}

\begin{figure}
\begin{center}
\begin{tabular}{c}
\includegraphics[width=0.45\textwidth]{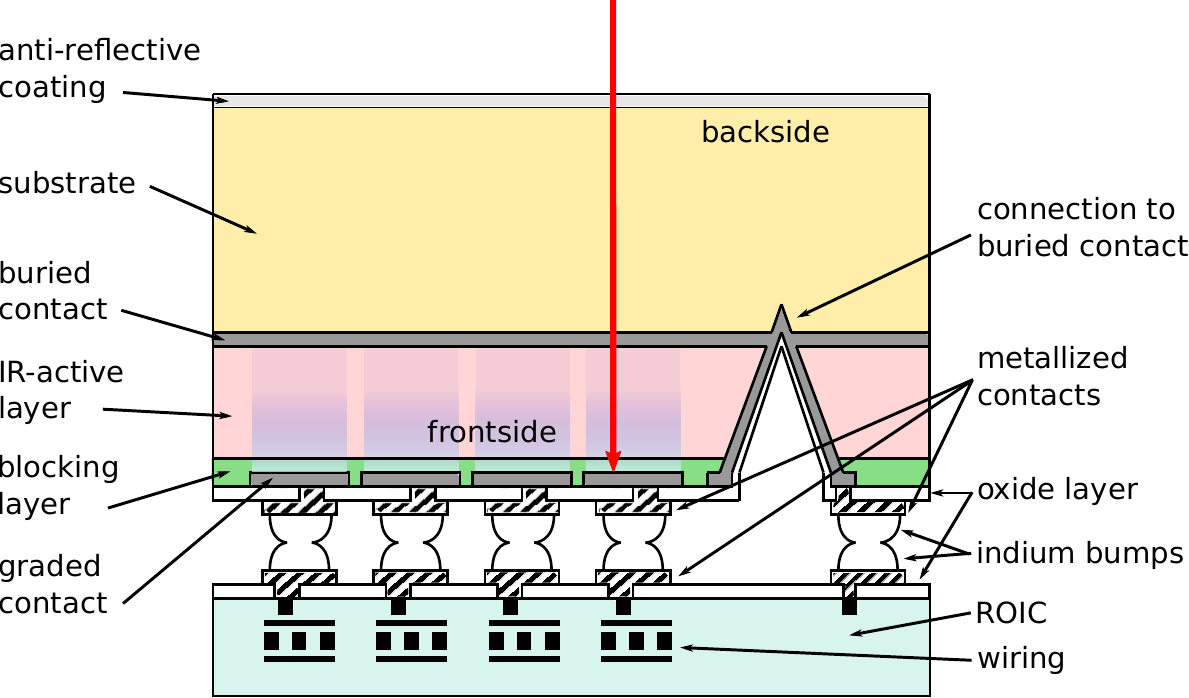}
\end{tabular}
\end{center}
\caption
{ \label{fig:xsection}  Cross-sectional diagram of the general architecture of a MIRI detector array (not to scale). Four pixels are shown with contacts and indium bumps to allow connection to their amplifiers in the readout integrated circuit (ROIC). The detector bias is established between the output contacts and the buried one. The individual electric fields of the four pixels are shown with a transparent blue gradient. The detectors are illuminated through their “backside,” i.e., through the substrate that supports them, rather than directly into the detectors, i.e., the “frontside.” The typical path of  photons is shown in red: they undergo reflective losses at the entrance and reflective and absorptive losses at the buried contact. If they are then absorbed in the IR-active layer, they create a photoelectron that traverses this layer and the blocking layer to be collected at a contact.  }

\label{xsection}
\end{figure}

The architecture of the detectors themselves is illustrated in Figure~\ref{fig:xsection}.  They are built on a silicon substrate about 500 $\mu$m thick.  The photon absorption occurs in an infrared-active layer, doped with arsenic, and 35 $\mu$m thick (baseline material), carefully cleared through wafer processing of minority impurities. A thin ($\sim$ 4 $\mu$m) high purity blocking layer intercepts thermally generated free charges in the impurity band that would otherwise produce  dark current and only lets photoelectrons in the conduction band of the material pass through to the amplifier. A bias voltage is established between the frontside contacts and a buried contact that separates the infrared-active layer from the rest of the substrate. The field established by this voltage depletes the infrared active layer of thermally released charge carriers, raising its impedance to very high levels. This field also drives photoelectrons from their creation sites to the frontside contact for the appropriate pixel. There they are conveyed through an indium bump to an output amplifier. 

\begin{figure*}
\begin{center}
\begin{tabular}{c}
\includegraphics[width=0.98\textwidth]{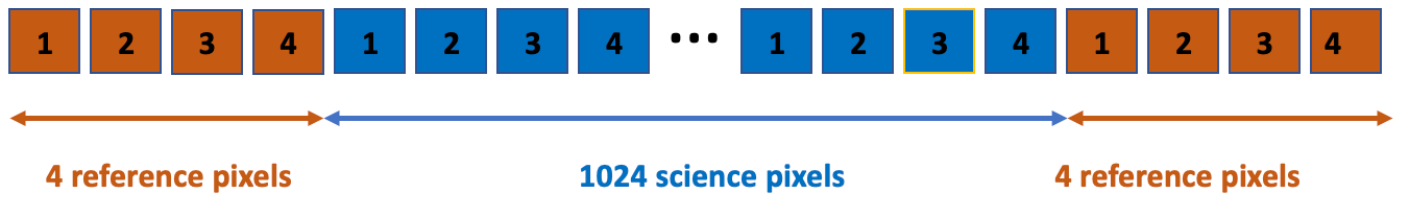}
\end{tabular}
\end{center}
\caption 
{ \label{fig:miri_pixel_row}
Diagram of the readout for one row with the science data with the reference pixels at the edges. The numbers indicate the output used to read out the pixel. } 
\end{figure*}

The MIRI SCA (sensor chip assembly) has a total of 1024 X 1024 active pixels. There are four additional reference pixels at the beginning of each row and four at the end that are unconnected "pixels" with no light sensitivity, but in all other ways are treated as light sensitive pixels. Figure  \ref{fig:miri_pixel_row}  is a schematic of this design for a row.  Including these reference pixels, the entire science image has a size of 1032 X 1024 pixels. All these pixels (including the reference pixels) are read out through four data amplifiers; each amplifier presents 258 X 1024 pixels to the signal chain electronics for processing. There is an additional "reference output" that is essentially a clocked DC voltage source; this may be used for eliminating noise. This reference output is read out with a fifth output line. The readout of the reference  line occurs simultaneously with the other four readout lines. 

\subsection{Readout Process}
MIRI employs JWST’s standardized readout of up-the-ramp (MULTIACCUM) sampling. To maintain the stability of the sensor chip assembly, the MIRI detectors are clocked out at a constant rate whether observing or not. The pixels are continuously addressed at time intervals of 10 micro-seconds and read out non-destructively. The standard terminology for MIRI is to define a 'group' as a single, non-destructive read of all the pixels in the array. An integration is defined as the time between the reset of the pixels (destructive read). There is one extra reset after each integration to help clean out  charge remaining on the readout after the reset . Each integration results in a ramp that, when fit, yields a measure of the flux per pixel (DN/second). One exposure consists of one or more integrations, each of which contains a number of groups.

\begin{figure}
\begin{center}
\begin{tabular}{c}
\includegraphics[width=0.45\textwidth]{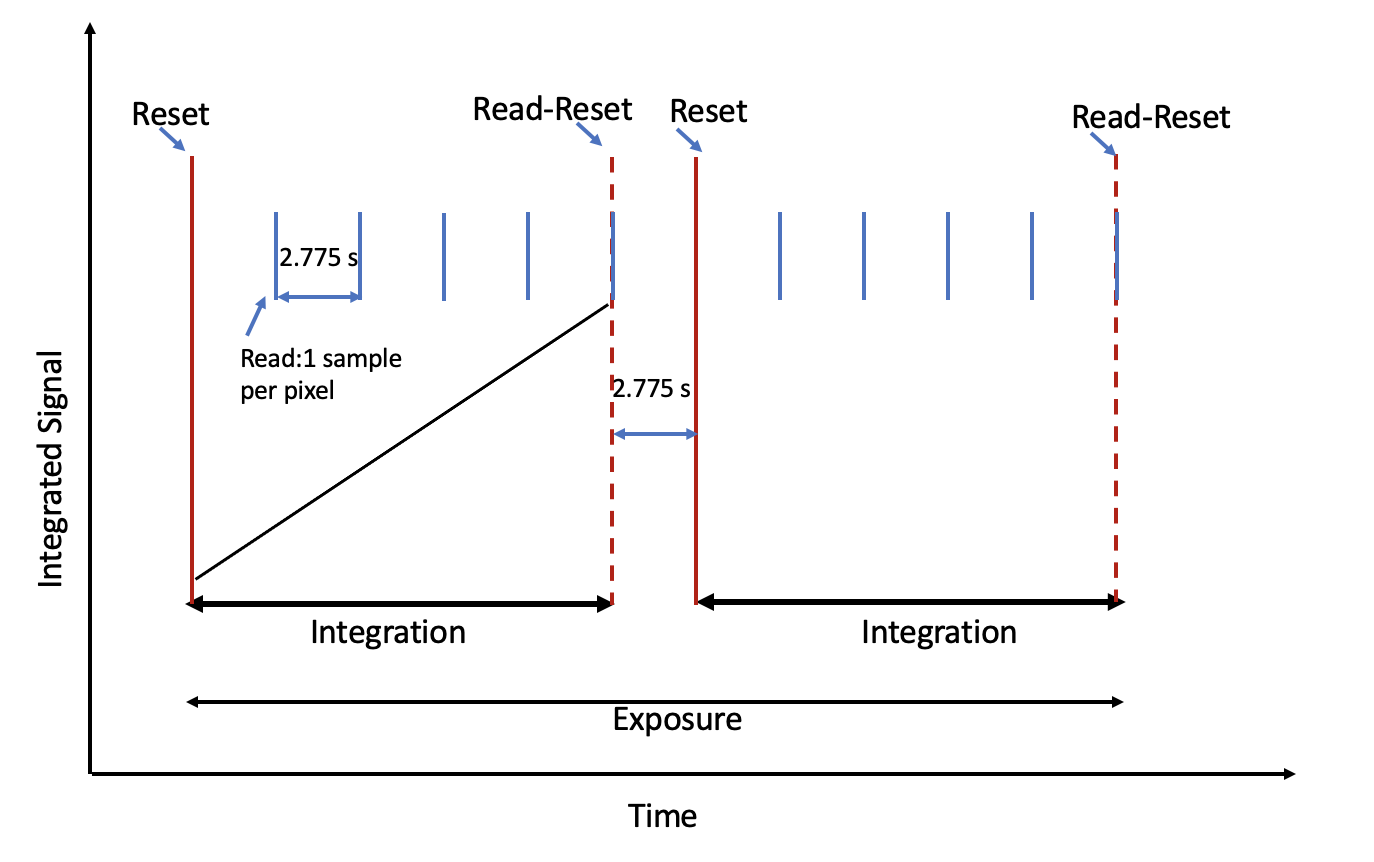}
\end{tabular}
\end{center}
\caption 
{ \label{fig:readout_FASTR1_v2}
The sampling up-the-ramp for the FASTR1 mode readout scheme for a single pixel. In this example  Nsample = 1, Ngroups = 5, and Nint = 2. The x-axis is time and the y-axis is signal strength. Each blue vertical tick marks a read of the pixel and each red vertical tick marks a reset of the pixel. In this scheme, each reading of a pixel will have only one sample at each pixel. The time to read an entire group is 2.775 s. } 
\end{figure}

There are two readout modes for MIRI: FASTR1 and SLOWR1. In FASTR1 mode the pixel is sampled once during a single clock cycle spent on that pixel. Figure~\ref{fig:readout_FASTR1_v2} is a schematic of the readout for a single pixel in FASTR1 mode for an integration with five groups and two integrations. The SLOWR1 mode samples the pixel 8 times within a 9 sample wide window; the first sample is ignored and the remaining 8 samples are averaged to output a single result.  The header keyword, NSAMPLE, is the number of samples per pixel, which will always be 1 for FASTR1 data and 9 for SLOWR1 data. This is a fixed parameter and cannot be changed by the observer. Because the entire sensor chip assembly is read before returning to the pixel, the time between  sampling an individual pixel is 2.775~s for FASTR1 and 23.890~s for SLOWR1. The final group of the integration is a read-reset, yielding the total of Ngroups reads. In multiple integration data the read-reset is followed by an additional reset frame that takes 2.775~s for FASTR1 and 23.890~s for SLOWR1. Because of the time difference in the pixel sampling between FASTR1 and SLOWR1 mode data, pipeline steps and reference files that contain time dependent information will have different calibration files depending on the MODE of the data. This is also true for different clocking patterns that occur in reading out the various subarray data.
 
The reference output data are rearranged into a 1032 X 256 image that is stored in the REFOUT extension of the uncalibrated science file. The general readout scheme is: (1)  the read out starts with addressing row 1 and then reading all 1032 columns; (2) after the first row is completed, it proceeds to the second row; (3) the two rows are reset simultaneously. This pattern is repeated until row 1024 is completed, with pixels always reset by row pairs. Our definition of even and odd rows throughout this paper is based on counting the first row on a detector as row 1 as an odd row.  Odd row data numbers (DN)  are slightly shifted with respect to even row data numbers. This shift is caused by the resetting of the row pairs and is also seen in the reference pixels. Therefore, when using the reference pixels  to correct the science data this even/odd offset needs to be folded into the correction (i.e., only use even row reference pixels to correct even row science data  and correspondingly for odd row data.). In addition, several other detector effects show an even/odd row dependence. 

\subsection{Subarrays}
The Imager detector is the only MIRI detector with subarray modes. They are implemented  by reading out partial sets of pixels through the manipulation of the clocking patterns. This allows a reduction in the frame time, and thus will allow integration times of substantially less than 2.775~s. For all subarrays, the subarray portion (the region-of-interest) is read out and stored while the remaining rows of the array are continually reset.  There are nine in total: (1) one for each of the four coronagraphic areas; (2) one for high backgrounds; (3) three for bright objects; and (4) one for slitless low resolution spectrometer (LRS). 

\subsection{Detector Testing}
Si:As IBC devices are the highest performance detectors that cover the wavelength range of 5 to 28 $\mu$m and are suited for infrared space missions. They provide high quantum efficiencies, low dark currents and resistance to damage from the effect  of cosmic rays. However, despite their outstanding performance, operating these arrays at temperatures of only a few kelvin (6.4 k) is still challenging and requires careful calibration to improve the performance and remove artifacts. Extensive investigations using flight-clone electronics and flight-like detector arrays were performed at the NASA Jet Propulsion Lab \citep[JPL, ][]{ressler2015}. The detectors were mounted inside a cryo-vacuum module to put them at their operating temperature of 6.4K.  In addition, before the launch of the JWST telescope the MIRI instrument underwent several ground test campaigns, which also included detailed  characterization of the various detector effects. These tests were conducted at the Rutherford Appleton Laboratory (RAL) in the UK,  Goddard Space Flight Center and Johnson Space Flight center in the US. These tests revealed several areas that needed follow-up testing at JPL.  From this vast testing of the MIRI flight and engineering detectors, the extended MIRI instrument team (Space Telescope Science Institute, European Consortium, JPL, and the University of Arizona) determined an initial set of algorithms designed to mitigate the non-ideal detector effects. More importantly from the commissioning and Cycle 1 data gathered from flight data operating in a space environment,  we have been able to further improve the algorithms and determination of the reference files.

\section {Pipeline Processing of MIRI data}
Many of the non-ideal behaviors of the MIRI Si:As IBC detector arrays are removed in the JWST calibration pipeline \cite{jwst_pipeline2022}. This includes effects such as integration ramp non-linearity, reset effects, readout issues, and others. Taken all together, these issues pose a challenge to accurate calibration, since the full suite of issues is likely to be influenced by a large variety of parameters -- history of illumination, illumination level, time since last observation, and so forth. The approach the MIRI team has adopted, as far as possible, is to separate the issues in a way that corrections can be made for one of them at a time in a simple way with a minimum of free parameters.

\begin{figure}[tbp]
\epsscale{1.1}
\plotone{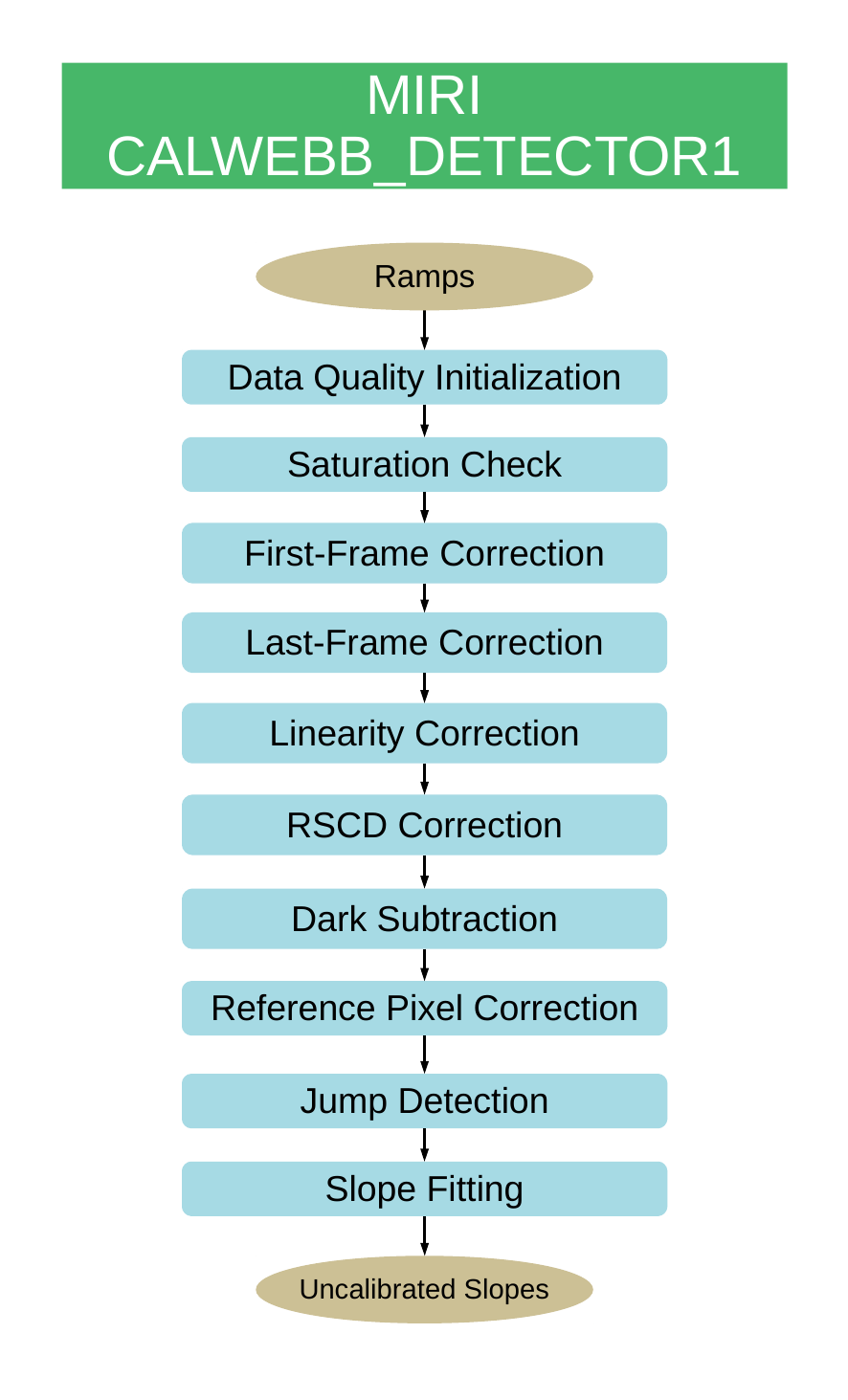}
\caption{The flow of the data reduction from the raw ramp data to uncalibrated count rates  (slopes) is given. 
\label{fig:flowchart}}
\end{figure}

We have researched and developed algorithms for the high priority issues. In this section of the paper we will give an overview of the effects that are corrected in  stage 1 of the JWST pipeline. The relevant steps are shown in the flow chart of  stage 1 of the pipeline, {\sc calwebb\_detector1},  as given in Figure~\ref{fig:flowchart}. Not all the MIRI non-ideal effects are corrected in the pipeline. In Section \ref{NoCorrection} we list some of the known effects we have not corrected and  will describe how, in some cases, these effects have impacted how the MIRI detectors are operated.

In this paper we will not discuss steps that are standard corrections in the JWST pipeline for all the JWST instruments; these include Data Quality Initialization, Saturation Check,  and Ramp Fitting.  For details on these steps and all the JWST algorithms see Gordon et al. (in preparation).

\subsection{First Frame effect}

\begin{figure}
\begin{center}
\begin{tabular}{c}
\includegraphics[width=0.45\textwidth]{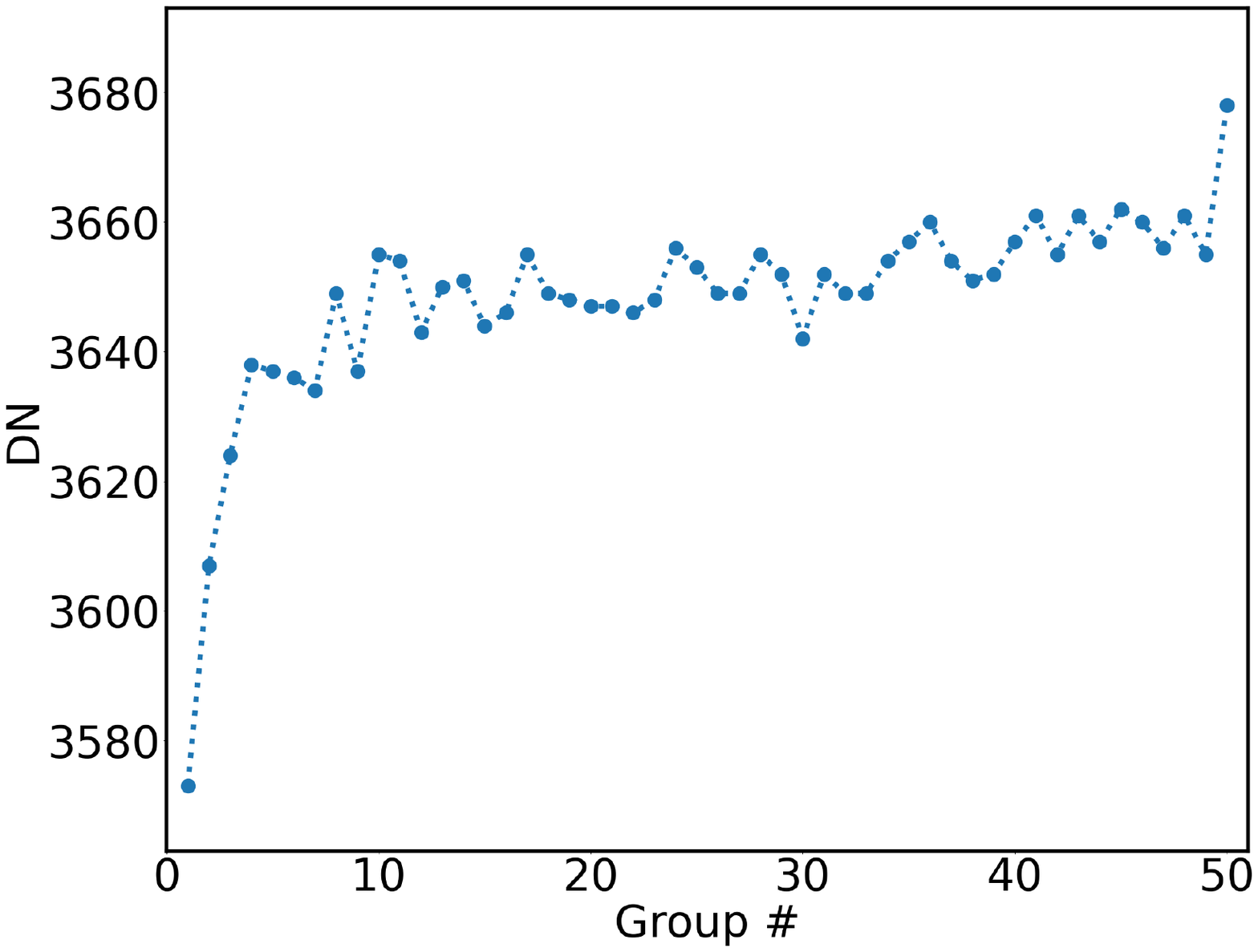}
\end{tabular}
\end{center}
\caption 
{ \label{fig:dark_ramp}  Pixel ramp for a dark exposure.  Features to note: the  non-linear region at the start of the integration and the last group is higher than expected. }
\end{figure}

As shown in Figure \ref{fig:dark_ramp},  the first few groups in an integration ramp are offset from their expected linear accumulation of signal. The majority of this deviation is related to the resets right before the start of the integration. We are able to correct for much of this deviation for groups two and higher; however a stable correction for the first group has not been found. The first group in an integration varies more than the other groups affected by reset. This variation depends on how long the detector has been in idle-reset (i.e. how long since the last MIRI exposure). 

\subsubsection{First Frame rejection in the JWST pipeline}
At this time we are unable to find an adequate calibration technique to correct the first group. Until a separate correction can be determined, the first group in an integration is being skipped by the JWST pipeline when fitting the slope of the ramp (i.e. determining the flux). In the near future we may relax this restriction for bright data with very few groups in an integration as this effect is small compared in this case.

\subsection{Last Frame Effect}
\begin{figure}
\begin{center}
\begin{tabular}{c}
\includegraphics[width=0.45\textwidth]{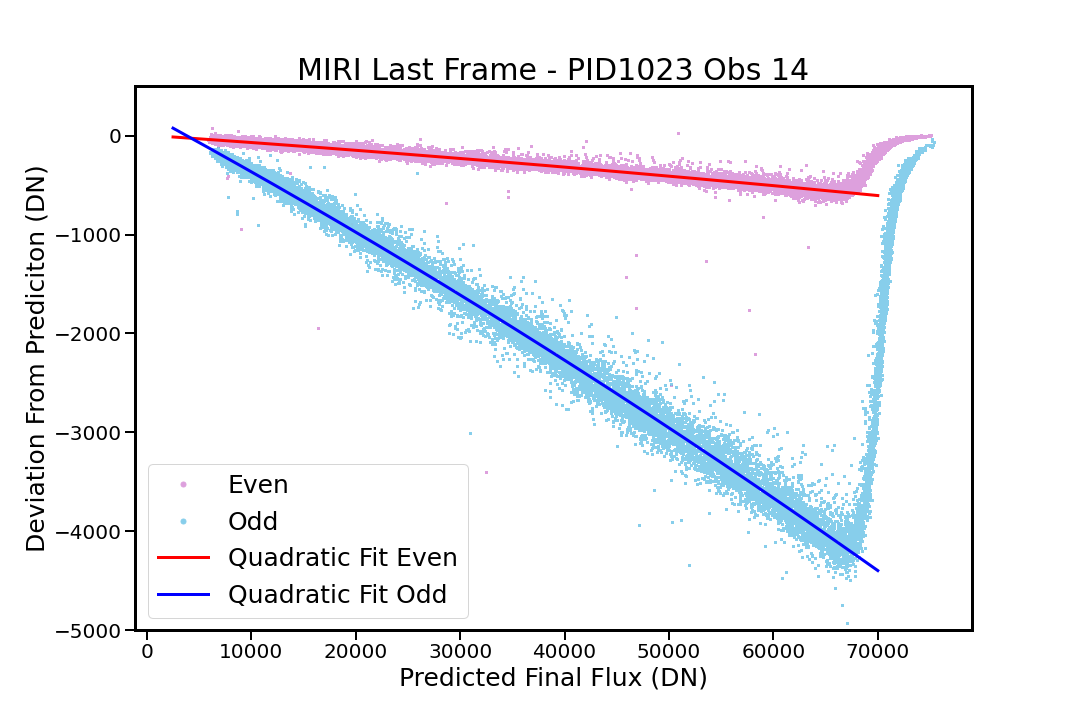}
\end{tabular}
\end{center}
\caption 
{ \label{fig:last_frame}
A plot showing the systematic offset in the last group of each integration from the predicted flux based on the prior count rate. Odd rows show an offset of approximately 6\%, while even rows show an offset of only about 1\%. The trendlines are given by a second degree polynomial to account for the slight non-linear shape of the effect, but are dominated by the linear term. As the final group approaches saturation, the effect goes away. } 
\end{figure}

Because the MIRI detectors are reset sequentially by row pairs,  the last group of an integration ramp on a given pixel is influenced by signal coupled through the reset of the adjacent row pair. The result is that the odd and even rows both show anomalous offsets in the last read on an integration ramp. The last group on dark exposures is always offset upward from the expected linear accumulation of signal (Figure \ref{fig:dark_ramp}) while on illuminated data the last group is offset downward and the amplitude of the downward offset is proportional to the signal in the previous group. Figure \ref{fig:last_frame} shows the deviation of the last group for even rows (purple) and odd rows (blue) from the predicted flux, based on the count rate from the previous group.

\subsubsection{Last Frame rejection in the JWST pipeline}
No adequate correction method has been found that corrects the last group and also increases the signal to noise of the fit of the ramp using this last group. Therefore, at this time, the baseline correction is to not use this group when performing the ramp fit.

\subsection{Electronic Non-linearity Correction}
\begin{figure*}
\begin{center}
\begin{tabular}{c}
\includegraphics[height=8cm]{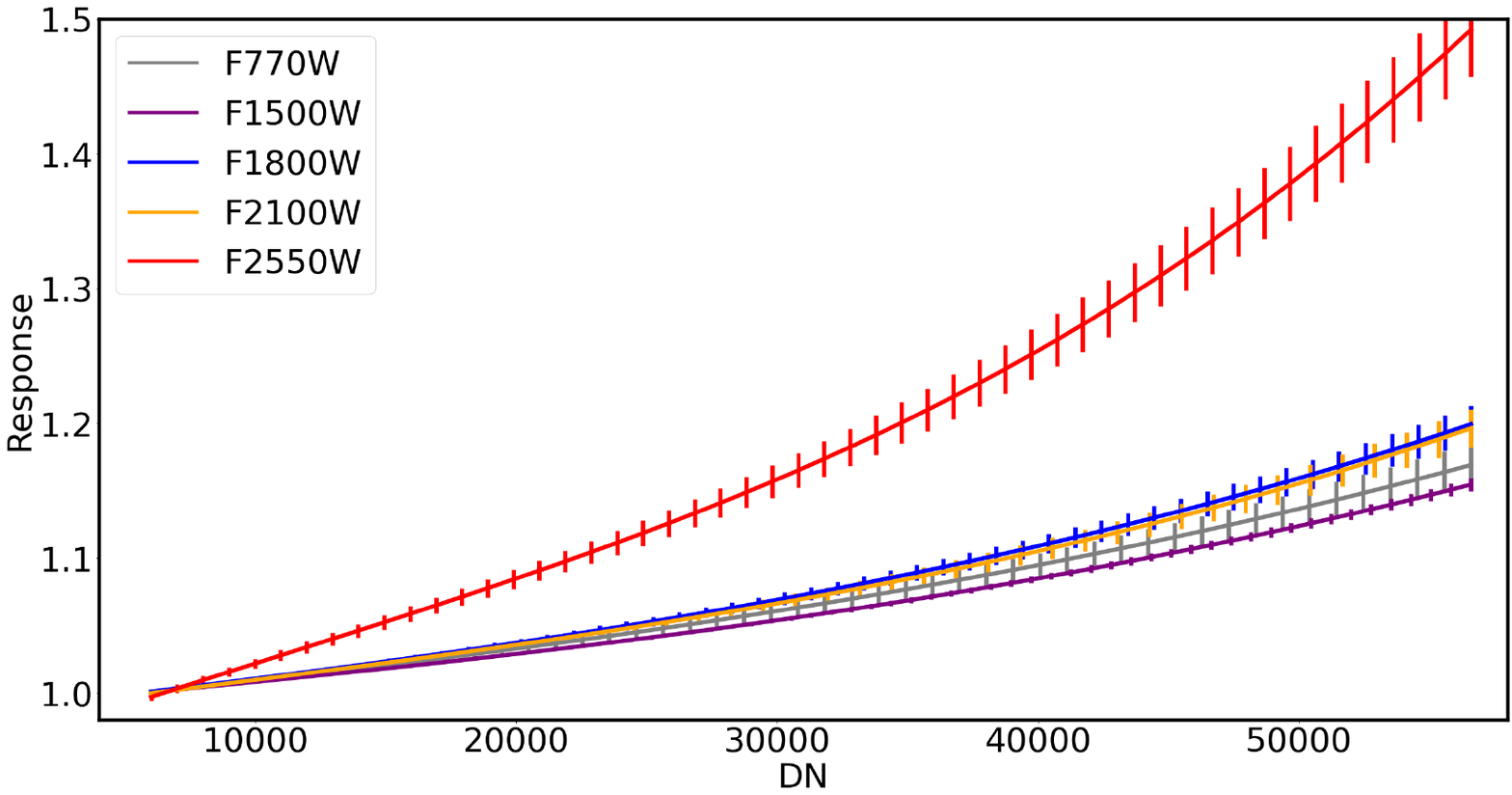}
\end{tabular}
\end{center}
\caption 
{ \label{fig:nl_imager}  Response curves and uncertainties for Imager detector. Here the response, y-axis, is the ratio of the a linear fit to the pixel ramps/ raw DN values of the pixel ramps. The wavelength-dependence of the non-linearity correction is evident from the divergence between the F1500W and the F2500W response curves.}
\end{figure*}

The MIRI detector outputs show  non-linearity in the measured integration ramps: there is a reduction in responsivity with increasing signal. This behavior arises because the detector bias voltage decreases as charge accumulates on the integrating nodes. As the bias is decreased, the detector depleted region shrinks toward the "blocking layer" (see Figure~\ref{xsection} and  \cite{Argyriou2023}) and the efficiency of collection of the photoelectrons decreases. The effect is strongest for $\sim$ 15 -- 26~$\mu$m, where the absorption is high and thus occurs primarily in the region farthest from the "blocking layer". In comparison, at the shorter wavelengths, e.g., 5 -- 10~$\mu$m, the absorption occurs throughout the IR-active layer, so the reduction of the detector bias and resulting shrinkage of the area from which the field collects free electrons has less effect on the net signal. The varying response curves for the Imager detector  are shown in  Figure \ref{fig:nl_imager}. 

Because of this difference in absorption for photons of different wavelengths and the resulting difference in photoelectron collection efficiency, the MIRI non-linearity correction is wavelength dependent. For the Imager detector, a correction derived using  data taken with the 7.7~$\mu$m filter is used for all the wavelengths at or below 12.80~$\mu$m. Separate corrections are used for each of the 15, 23 and 25.5~$\mu$m filters; while data from the 18 and 21~$\mu$m filters use a correction derived using the 21~$\mu$m filter.  For the MRS long wavelength detector (11.55 -- 27.9~$\mu$m) there is a separate correction for each spectral band. However, just one correction is used for all the spectral bands of the MRS short wavelength detector (4.9 -- 11.7~$\mu$m).   
 
 The behavior of the electronic non-linearity was determined first through considerable ground-based testing on the engineering and flight arrays. The MIRI instrument has  two on-board calibration sources mounted in the MIRI Optics Box Assembly. They provide stable and smooth (homogeneous) illumination of the Spectrometer and Imager optical paths. The design uses two identical Calibration Hot Source Units, one in the Imager Calibration Unit  and the other in the Spectrometer Calibration Unit.
 During ground testing the calibration lamps were used to study the characteristics  of the MIRI electronic non-linearities. Exhaustive studies  revealed that there were insignificant differences in the non-linearity between pixels, and a single global solution could be used to correct  the pixel ramps for  electronic non-linearities to less than 0.25\% deviation from  linear behavior  on average.
 
 In commissioning on orbit, the calibration lamps were used to obtain well illuminated pixel ramps up to saturation both for the majority of the pixels for the Imager detector (using the 7.7~$\mu$m filter) and for all three spectral bands of the MRS short and long wavelength detectors.  During Cycle 1 calibration, similar data were obtained for the 25.5~$\mu$m, 21~$\mu $m, 18~$\mu$m, and 15~$\mu$m filters using a sparsely populated region on the sky\footnote{The non-linearity correction was derived using pixels with no detectable sources.}.  There is ongoing work on obtaining data for the 23~$\mu$m coronagraphy filter.
 
 The technique to derive the linearity correction used a grid of 16 $\times$ 16 points evenly spaced across the detector. At each grid point a 6 $\times$ 6 pixel box was used to select
 pixels with approximately the same signal, and after screening for cosmic ray hits and bad pixels the remaining pixels at the selected grid point were used to derive a mean ramp with empirically derived standard deviations. After removing the dependence of the varying signal across the 16 $\times$ 16 grid points, either from the calibration lamp or the background,  global fits were used to derive a  set of linearity coefficients to use for all the pixels on the detector. The non-linearity for most of the pixels is adequately described as a cubic polynomial. The linear component of this fit gives the linearized signal.
 In addition, as will be discussed in Section \ref{bfe}, the  contrast of the MRS spectral slices adds another layer of complexity to the derivation of an adequate non-linearity correction. Currently pixels on fringe peaks were selected to minimize the brighter-fatter effect (described in Section \ref{bfe}) and \cite{Argyriou2023b} to derive a non-linearity correction. Future improvements in the MRS electronic non-linearity correction will need to fold in the brighter-fatter effect \cite{Argyriou2023b}.

\begin{figure*}
\begin{center}
\begin{tabular}{c}
\includegraphics[height=18cm]{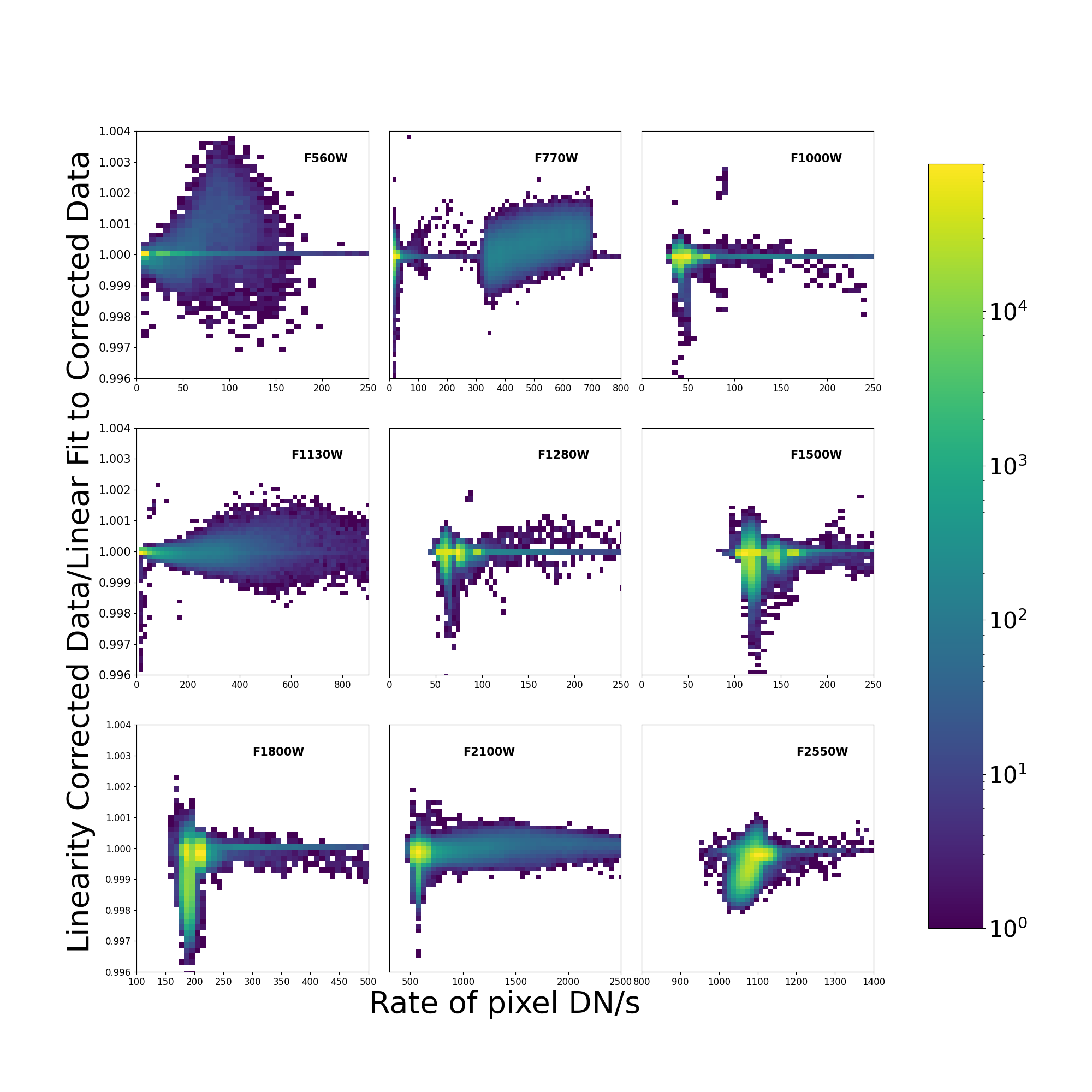}
\end{tabular}
\end{center}
\caption 
{ \label{fig:flux_linearity} 
An analysis using 812 exposures from the Imager detector, testing the quality of the linearity correction. This 
 2-D histogram bins the data by pixel rate (~uncalibrated flux) and the  pixel linearity metric (mean ratio between the linearized data and a linear fit of the linearity corrected data). The data was split by filter to check for any wavelength dependence in the linearity correction, and the panels show the results for the  F560W, F1000W, F1130W, F1280W, F1500W, F1800W, F2100W, and F2550W filters.  The color for each bin  (as coded by the color bar on the right) represents of the number pixels in the linearity ratio and rate bin. This  plot shows there is no flux dependence to the Imager linearity corrections and the correction for the majority of the pixels corrects the ramps to better than 0.02\% linear.}
\end{figure*}

We evaluated the linearity correction using 812 Imager exposures from commissioning and cycle 1 with at least five groups per integration and selected to have wide ranges of fluxes on the images. These exposures were processed with the JWST pipeline with the option to write out the linearity corrected ramp. Pixels flagged as having a jump in the ramp (i.e., a cosmic ray hit) were not used in the test. To avoid issues related to the reset switch charge decay effect (RSCD, see Section \ref{RSCD}) we only used the first integration in this test.  For full frame data we used pixels in a 400 X 400 pixel region in the center of the Imager field of view ($500 < x < 900$ and $500 <  y < 900$). Our test set included a sampling of all the subarrays except for SLITLESSPRISM. For the subarray data we used a region centered on the subarray and having a size 1/2 the corresponding x and y axis.  In addition, to focus on pixels with higher flux, where the linearity correction is important,  we limited our analysis to pixels with a rate value greater than 5 DN/s. For these pixel ramps,  we fit a polynomial of degree 1 to the linearity corrected group values which do not saturate.  For each pixel we saved the slope (rate)  from the linear fit and  found the ratio between the linearity corrected group value and the group value predicted from the linear fit.  Next for each pixel we determined the mean of this ratio (we will call this the mean linearity ratio). This ratio is one metric we use to determine how well the linearity step has linearized the pixel ramps. A value equal to 1 shows linearization is achieved, a value greater than 1 indicates the ramp values are over-corrected in the linearity step, and a value less than 1 shows an under-correction in the step. We split the data by filter to check the wavelength dependence of the linearity correction. To  test if there was flux dependence in the linearity correction, we created a 2-D histogram of the  mean linearity ratios to the pixel rates  (which  represents the uncalibrated flux). Fig.~\ref{fig:flux_linearity}  is a  2-D-histogram of  linearity ratio vs the pixel rates.  We plotted the results for filters ranging from  F560W to F2550W filters. The color for each bin (as coded by the color bar on the right) represents the number of pixels in each linearity ratio and rate bin. This  plot shows there is no flux dependence to the Imager linearity corrections and the correction for the majority of the pixels corrects the ramps to better than 0.02\% linear.

\subsubsection{MIRI non-linearity correction in JWST pipeline}
Through a careful study it has been found that the linearity correction should be applied to measured value data numbers (DN) (for the NIR detectors the correction is applied to the bias subtracted DN level). For this reason there is no superbias subtraction step in the {\sc calwebb\_detector1} pipeline for the MIRI detectors.  The goal of the non-linearity correction is to  adjust the integration ramps so that the output of the adjusted DN is a linear function of the input signal. At this time, based on the evaluation of nonlinearity over the area of the detector arrays, a single correction is used for all the pixels on the detector in a given wavelength range. This assumes that the non-linearity of the detector is constant across the detector.

\subsection{Reset Effects}
\subsubsection{Cause of Reset Effects}

\begin{figure}
\begin{center}
\begin{tabular}{c}
\includegraphics[height=8.0cm]{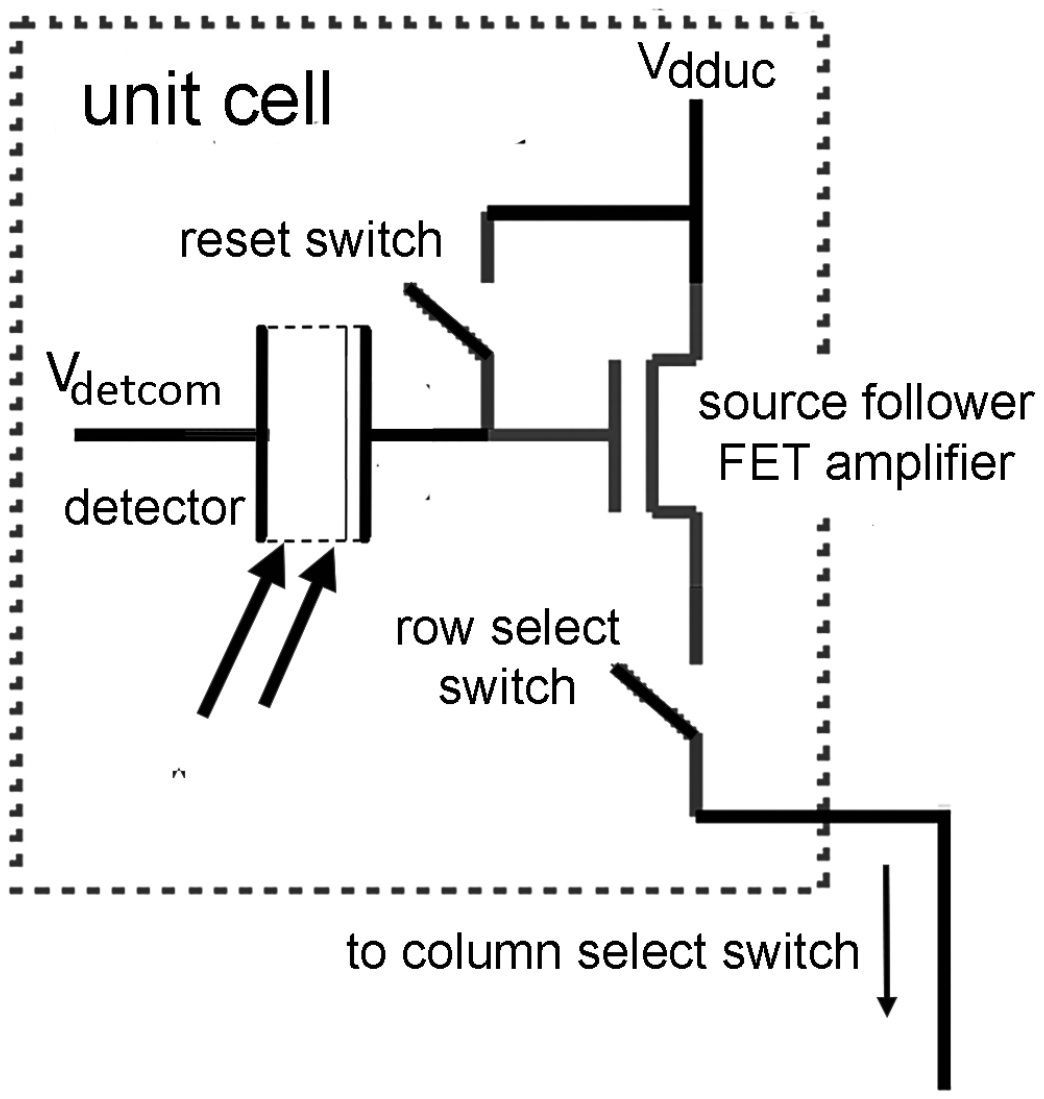}
\end{tabular}
\end{center}
\caption 
{ \label{fig:single_cell_roic}  Single cell of the MIRI schematic of the readout circuit.
}
\end{figure}

To explain effects seen in MIRI data that are related to resetting the detectors we need to expand on Figure~\ref{fig:xsection} and explain how the Read-Out Integrated Circuit (ROIC) in the sensor-chip assembly (SCA) works.  Below we give a short summary of the readout process \citep[for more details see ][]{ressler2015,laine2023}.  A simple schematic of a unit cell of read-out circuit is shown in  Figure~\ref{fig:single_cell_roic}. An initial  bias  voltage is set by closing the reset field-effect transistor (FET) switch to connect the \(V_{dduc}\) supply to the amplifier input and the detector frontside contact.  The detector bias voltage is set by the difference between \(V_{detcom}\) (applied to the transparent buried backside contact) and \(V_{dduc}\) (applied to the frontside  contact).  Closing the reset FET therefore  initializes the detector pixel and prepares it for integration. During an integration the reset switch FET is opened (i.e., the FET is off), and the detector bias voltage decreases as photons are converted to electrons, which are collected at the integrating node. At the end of the integration period, the final node voltage is read. The signal recorded by a pixel is the difference between this final node voltage and the initial bias voltage. 

In practice  there are a number of non-ideal reset effects when a FET is used as a reset switch in this type of circuit \citep{yu2010}. We will group them together and use the descriptive term \textit{Reset Switch Charge Decay (RSCD) effect}. This effect, as observed in infrared detectors, is a product of their low temperatures of operation and the resulting large effective resistances and long RC time constants. Specifically for the reset circuit, when  the reset FET is turned  “off,” the space charge in its channel relaxes slowly because it is isolated by high resistances. The relaxation appears as charge injected from the FET channel into the integrating node. Although the reset nominally sets the FET source and drain to the same voltage, after charge injection there is a voltage difference between them that drives a current between source and drain, allowing the integrating node to recharge toward the reset voltage. While this is occurring, the detector bias and responsivity are varying, and thus the integration slope is distorted. Once equilibrium has been achieved, the bias is fully restored and the integration ramp follows the equilibrium behavior. The voltage error introduced by the above process is proportional to the difference between the reset voltage and the source voltage; the larger the voltage difference, the larger the error. This type of error results in zero point shifts between integrations and an offset in the initial groups in an integration. The amplitude of both effects depends on the signal level in the previous integration. 

\subsubsection{RSCD effect in the ramps }\label{RSCD}

\begin{figure*}
\begin{center}
\begin{tabular}{c}
\includegraphics[width=0.95\textwidth]{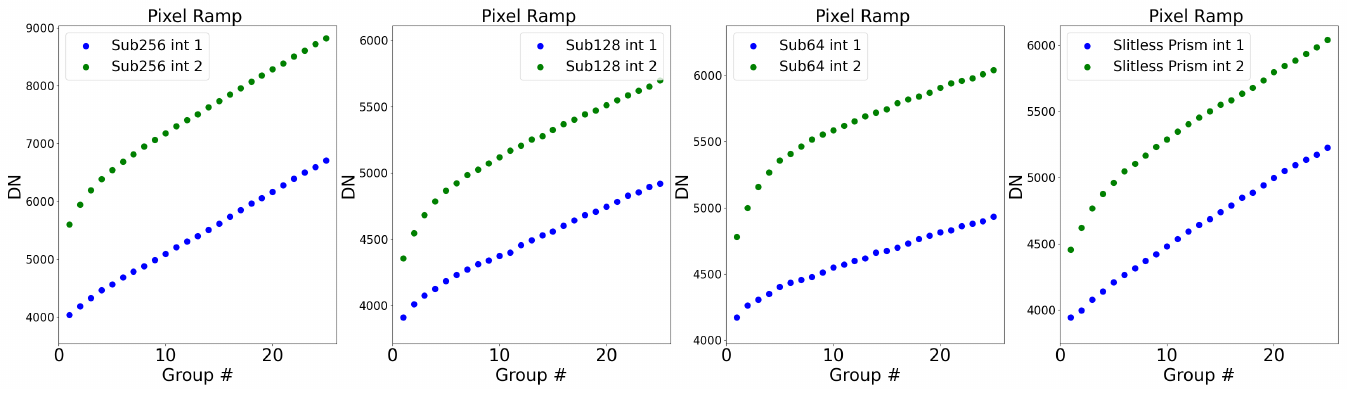}
\end{tabular}
\end{center}
\caption 
{ \label{fig:sub_rscd}  
The RSCD effects are illustrated for subarray exposures with two integrations.  The blue points are the first 25 groups of the pixel ramp for the first integration, while the green points are the same pixel ramp for the second integration.    The plots from left to right are pixel ramps for the SUB256, SUB128, SUB64, and SLITLESSPRISM subarrays}
\end{figure*}

The MIRI detectors are clocked out at a constant rate whether observing or not. Therefore, between exposures the detectors are read out and reset continually until the start of the next exposure. If an exposure has more than one integration, then there are more resets before the first integration compared to other integrations.  The difference in the number of resets before an integration results in a difference in the amplitude and profile of the RSCD effect which in turn results in different count rate for the same pixel from  different integrations.  Figure~\ref{fig:sub_rscd} shows a pixel ramp for four different subarrays. The pixel ramps for integration 1 (blue) and integration 2 (green) are shown (integrations 3 and higher are very similar to integration 2). 
The major components of the RSCD effects  illustrated in these figures  are:
\begin{itemize}
    \item \textbf{Fast Decay}: The second and higher integration ramps show an extra signal at the beginning of a ramp that decays exponentially. The first integration also exhibits this decay, but at a smaller amplitude because the preceding detector idle operation keeps the accumulated signals small. 
    \item \textbf{Slope difference}: The first and second integration count rates measured from the ramps after correcting for the known non-linearity differ. The  rate difference depends on the length of the integration (shorter integrations often show larger differences) and the read-out mode (subarray data generally shows a large effect than full array data.)
\end{itemize} 

\subsubsection{RSCD step in the JWST pipeline}
Ground testing at JPL showed that trying to model the varying amplitude and time scales of the RSCD effects was extremely difficult. The current RSCD step in the JWST pipeline does not try to correct for the fast decay of the groups at the start of an integration. Instead it flags initial groups in integrations two and higher to be skipped in the ramp fitting step. The number of groups rejected is based on the mode and subarray size and is stored in the RSCD reference file.  See Table \ref{table:rscd} for the number of groups rejected for the version 9.1 of the JWST Pipeline. The MIRI JPL team is studying an algorithm to correct for the fast decay which could be used in a future version of the pipeline \citep{laine2023}. 

\begin{table*}
\begin{center}
\begin{tabular}{|c|c|c|c|}
\hline
Detector & Subarray & Mode & Number of Groups to skip \\
\hline 
IMAGER & FULL & SLOWR1 & 0 \\
\hline
IMAGER & FULL & FASTR1 & 2 \\
\hline
IMAGER & BRIGHTSKY & FASTR1 & 2 \\
\hline
IMAGER & SUB256 & FASTR1 & 4 \\
\hline
IMAGER & SUB128 & FASTR1 & 6 \\
\hline
IMAGER & SUB64 & FASTR1 & 7 \\
\hline
IMAGER & MASK1550 & FASTR1 & 4 \\
\hline
IMAGER & MASK1140 & FASTR1 & 4 \\
\hline
IMAGER & MASK1065 & FASTR1 & 4 \\
\hline
IMAGER & MASKLYOT & FASTR1 & 4 \\
\hline
IMAGER & SLITLESSPRISM & FASTR1 & 4 \\
\hline
MRSLONG & FULL & SLOWR1 & 0 \\
\hline
MRSLONG & FULL& FASTR1 & 2 \\
\hline
MRSSHORT & FULL & SLOWR1 & 0 \\
\hline
MRSSHORT& FULL& FASTR1 & 2 \\
\hline
\end{tabular}
\caption{Number of groups to skip  at the start of an integration for integrations two and higher, due to the RSCD.}
\label{table:rscd}
\end{center}
\end{table*}

\subsection{Reset Anomaly and Dark Subtraction}
\begin{figure}
\begin{center}
\begin{tabular}{c}
\includegraphics[width=0.45\textwidth]{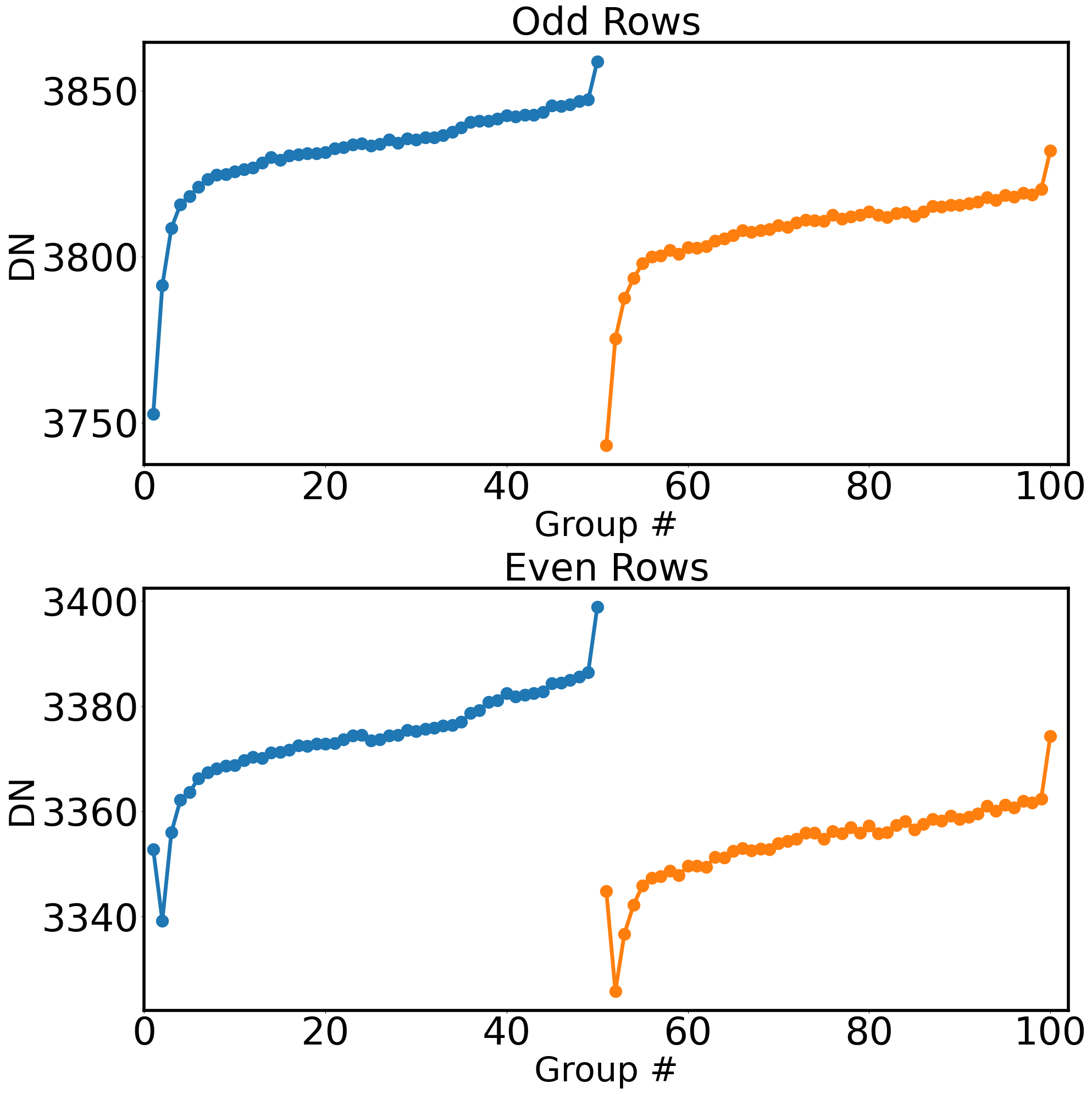}
\end{tabular}
\end{center}
\caption 
{ \label{fig:dark_odd_even} Top:  Mean ramp of neighboring odd pixels for the first (blue points) and second  (orange points)  integration. Bottom: Mean ramp of neighboring even pixels for the first and second integration.  200 pixel ramps were averaged to make the mean ramp, with sigma clipping to remove the effects of cosmic rays. }
\end{figure}

Dark exposures also show the effects of the reset in their ramps.  An example of a dark ramp showing the decay of the charge left on the reset FET from the dark current is shown in Figure \ref{fig:dark_odd_even}. These are plots of full array dark pixel ramps consisting of 50 groups ( $ \simeq 138 $ seconds) and two integrations. What is apparent in these plots is that the initial groups are offset from their expected values; there is a difference between even and odd rows, as well as between integration one and integration two.   

\begin{figure*}
\begin{center}
\begin{tabular}{c}
\includegraphics[width=0.95\textwidth]{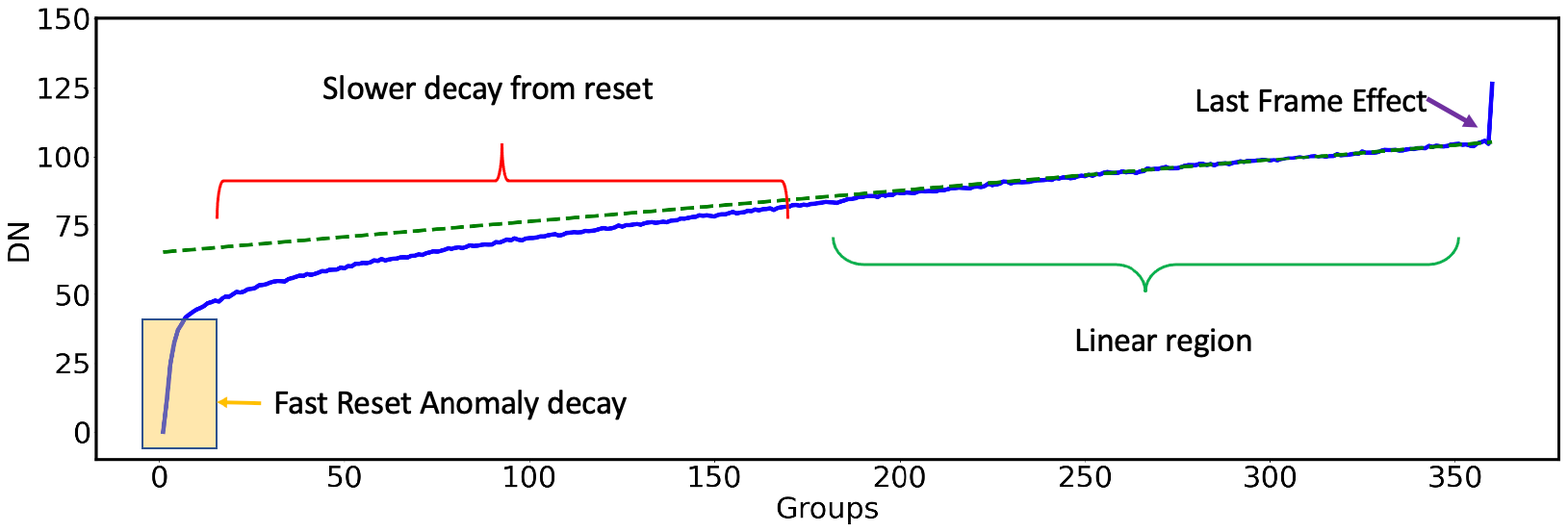}
\end{tabular}
\end{center}
\caption 
{ \label{fig:dark_exposure} Plot of a long dark exposure. The first group has been subtracted from each pixel ramp so that each ramp starts at 0. The blue line is the median of all the pixels at each group after sigma clipping for outliers caused by hot pixels, dead pixels and cosmic ray effects. The accumulated signal is not linear with time. The dashed green line is a linear fit to the last 150 groups.}
\end{figure*}

Figure \ref{fig:dark_exposure} is a global overview of what happens in a dark exposure. It illustrates a long full array dark exposure consisting of 360 groups ($ \simeq 1000$ seconds); we  subtracted the first group from all the pixels, and determined a group median after sigma clipping to remove the effects of cosmic rays, hot and dead pixels. A linear fit using groups 150 to 360 is shown in green.  The first $\simeq 100 $ groups deviate from the linear fit.  The large deviation of the initial groups has a short time-scale decay of $\simeq 15 $ groups for full frame data. The MIRI team has termed the non-linear fast decaying effects in dark exposures as the  \textit{reset anomaly}.  There is a secondary lower level  effect that lasts for $\simeq 100 $ more groups before the groups become, more or less, linear  with time and represent the true dark current. 

\subsubsection{MIRI Dark Subtraction in JWST pipeline}
The majority of the reset effects found in dark exposures  and thus in all exposures are removed by creating the dark reference file made from a stack of long dark exposures.   The zero point of the integration for the same pixel from different exposures varies through the stack. To compensate for the differing zero points and for cosmic rays that cause jumps in the ramps, the dark reference file is constructed from sigma clipped group differences. The final dark reference contains a value of zero for the first group and all other groups are determined from the sum of the sigma clipped dark group differences. As shown by Figure \ref{fig:dark_odd_even}, the amplitude of the reset anomaly is different in the first integration of the dark from the other integrations because of the varying number of resets at the start of an exposure.  The final dark reference file contains a correction for the first integration and a correction for all the other integrations. 
Dark reference files are created for each readout mode (FASTR1 and SLOWR1) and for each subarray configuration.  Applying the dark reference file removes the majority of the dark current,  and reduces the the differences between integrations and even and odd rows caused by the reset.  

\begin{figure}
\begin{center}
\begin{tabular}{c}
\includegraphics[width=0.45\textwidth]{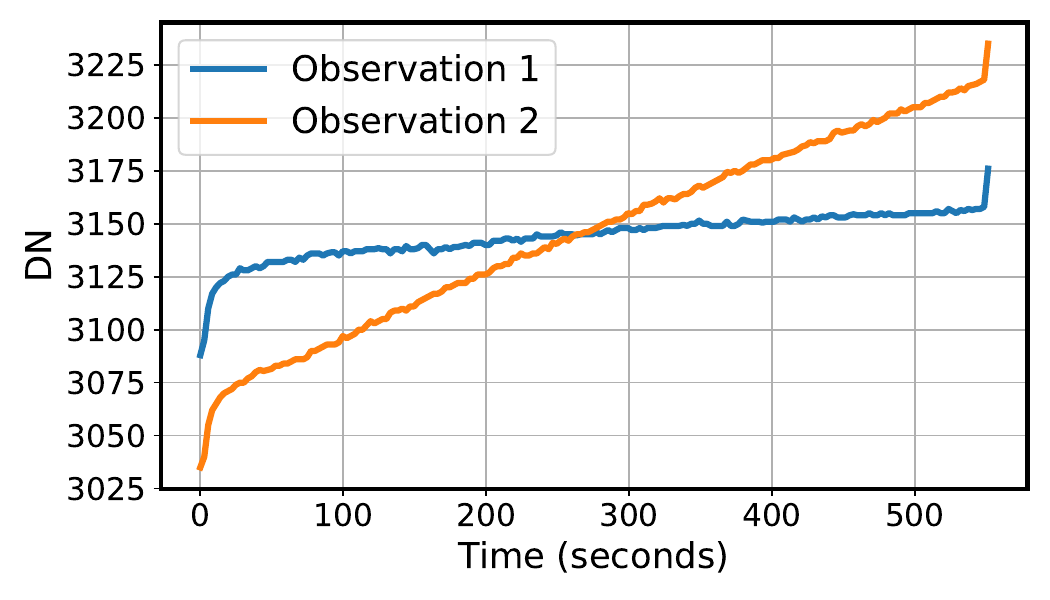}
\end{tabular}
\end{center}
\caption 
{ \label{fig:darks_vary} {Median MRS ramps in a roughly 6000-pixel dark region in the middle of the short-wavelength detector.  Observation 1 (orange line) and Observation 2 (blue line) are identically-configured observations taken two days apart, targeting shielded areas within already nearly-empty fields and shows  dramatically different effective dark count rates.}}
\end{figure}

\begin{figure*}
\begin{center}
\begin{tabular}{c}
\includegraphics[height=7.0cm]{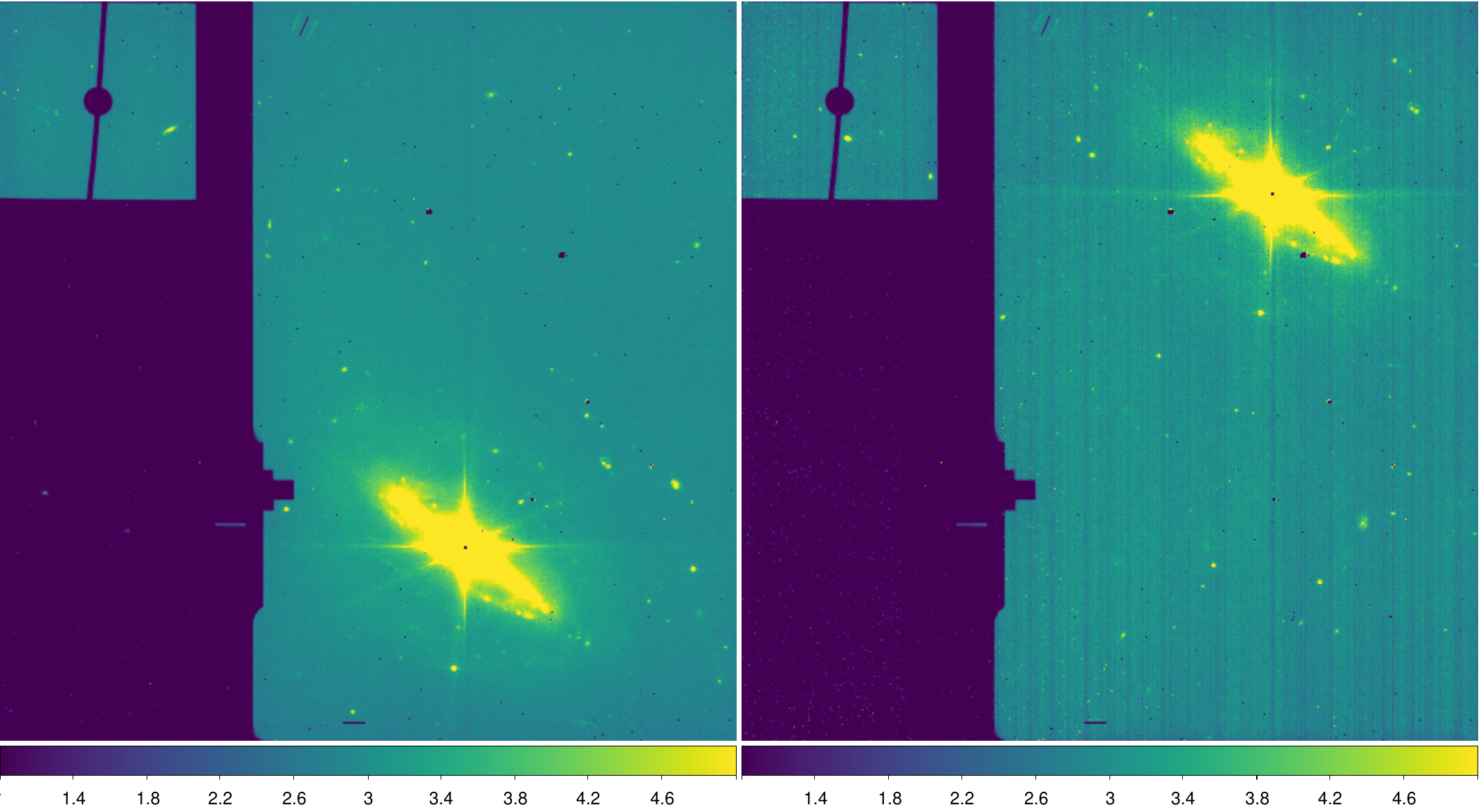}
\end{tabular}
\end{center}
\caption 
{ \label{fig:darks_structure} {Two images of the Seyfert galaxy NGC 6552 obtained with the F560W filter: the image on the left was observed with a single integration of 200 groups, while the image on the right was offset using APT and was observed with eight integrations of 25 groups. Both images have the same total integration time of 555 seconds.  }}
\end{figure*}

Unfortunately, the dark count rate appears to vary in time.  Figure \ref{fig:darks_vary} shows strikingly different count rates for two short wavelength MRS observations taken two days apart for pixels that do not receive light from the sky. Note the plot also shows the large last frame effect.  In addition, Figure \ref{fig:darks_structure} shows  the spatial variation in images that are predominately caused by the changing amplitude of the reset anomaly.
The image on the left was observed with a single integration of 200 groups, while the image on the right  taken 24 minutes after the first image was observed with eight integrations of 25 groups. Both images are from the same visit and have the same total integration time of 555 seconds.  The pipeline removes reset effects by subtracting a dark reference file. If the amplitude of the reset anomaly has changed from when the exposures used to create the dark reference file were taken, then  the pipeline will not be able to completely remove the reset effects. The reset effects are strongest in the initial groups  of an integration, and therefore the data with only 25 groups shows an incomplete reset anomaly correction.   We are currently exploring possible causes for the variable dark rate and algorithms to account for this variation. At present, the MRS {\sc calwebb\_spec2} pipeline applies a preliminary correction in the straylight step by simply subtracting the median residual dark rate measured in unilluminated regions between the two channels.

\subsection{Reference Pixel Correction}
\begin{figure}
\begin{center}
\begin{tabular}{c}
\includegraphics[height=7.0cm]{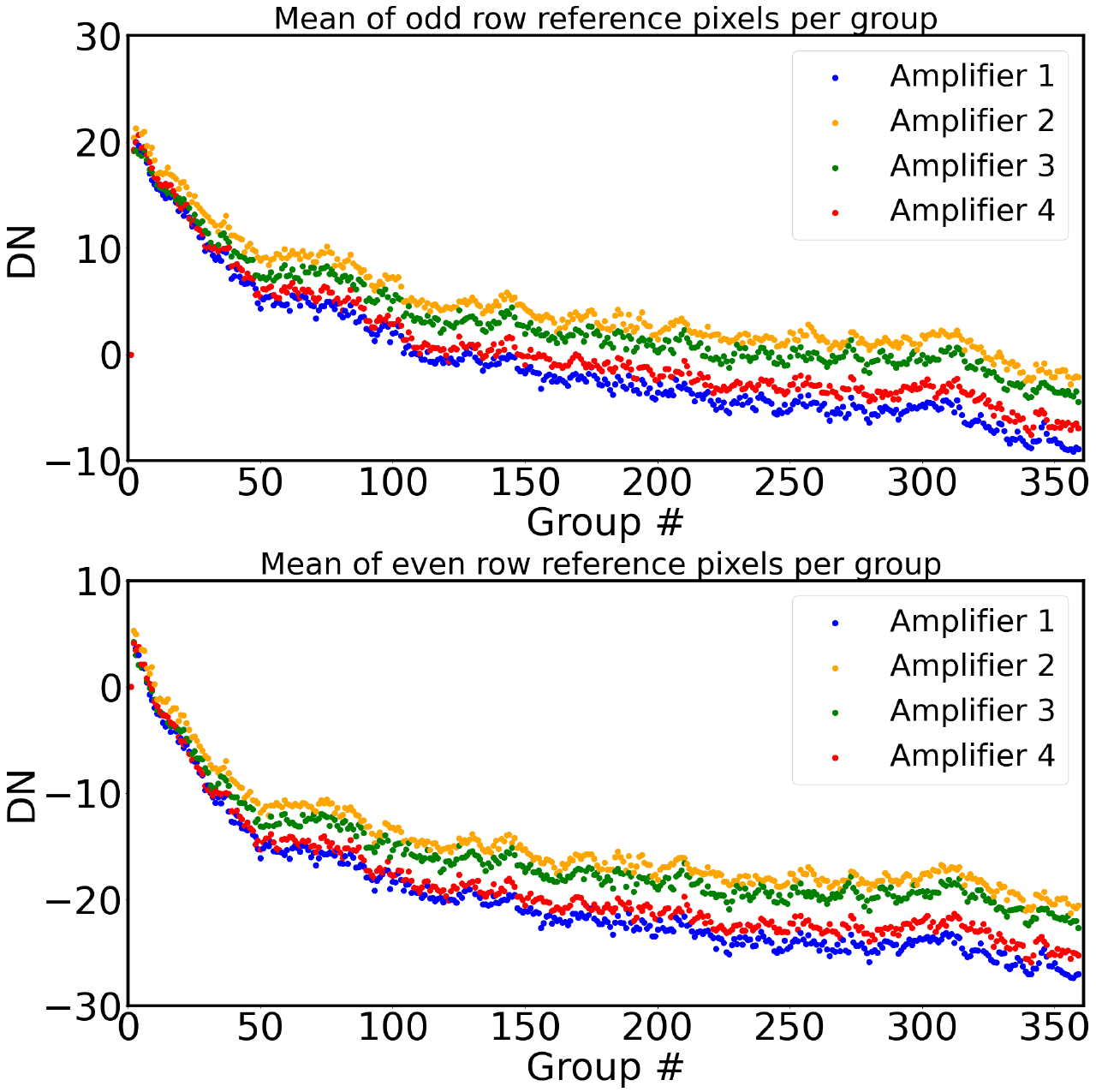}
\end{tabular}
\end{center}
\caption 
{ \label{fig:ref_pixels} {The average of the even and odd reference pixels for each amplifier from a dark  exposure with 360 groups. This mean is made after the first group was subtracted.  The mean reference pixel for group 1 for all the cases is zero. Over-plotting of the amplifier points obscures this mean of zero for amplifiers 2-4 }}
\end{figure}

The JWST infrared detectors are sampled multiple times non-destructively during an integration to derive a count rate for each pixel. A difficulty with this method is the observed slope is affected by drifts in analog reference voltages during the integration, leading to a correlated error for the pixels in the detectors. To address these electronic drifts, all the JWST science detectors implemented reference pixels. Recall from Figure~\ref{fig:miri_pixel_row} that there are four "reference pixels" at the beginning and end of  each row of the detector array (one on each side for each amplifier).  
Figure \ref{fig:ref_pixels} is a plot of the average reference pixel signals for each output (amplifier) for each group, after subtracting the first group to remove variations in the zero point of the reference pixel ramps across the detector. The mean of each group is determined separately for odd rows (top plot in Figure \ref{fig:ref_pixels}) and even rows (bottom plot).  There is a different profile for each amplifier, as well as for even and odd rows for a given amplifier. 

\begin{figure}
\begin{center}
\begin{tabular}{c}
\includegraphics[height=7.0cm]{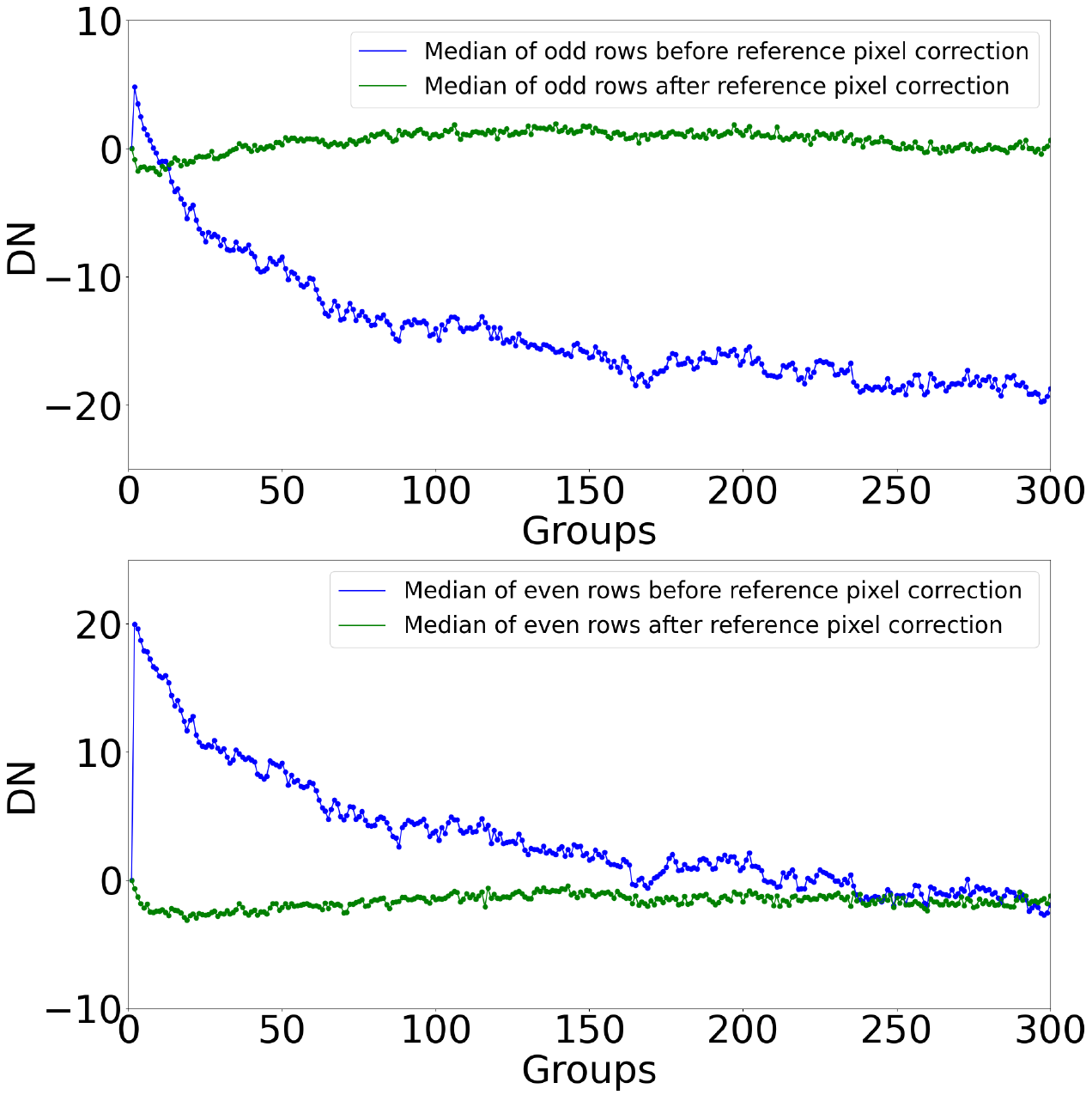}
\end{tabular}
\end{center}
\caption 
{ \label{fig:ref_pixel_correction} { The median  of all the pixels in a group in a detector for a dark exposure before the reference pixel correction (blue) and after it (green).  }}
\end{figure}

\subsubsection{MIRI reference pixel correction in JWST pipeline}
For full array data the reference pixels are used to remove noise due to drifts. The current algorithm subtracts the first group of a reference from all the groups in the reference pixel ramp.  Then for each group the reference pixels are separated by amplifier and whether they are on an odd row or even row. Taking the mean of each set results in eight mean values per group. These values are subtracted from the science pixels depending on the amplifier and whether a science pixel is in an odd or even row. 
Figure \ref{fig:ref_pixel_correction} is the median of each group across the detector before the reference pixel correction (blue points) and after the reference pixel correction (green points). The reference pixel correction has reduced significantly the electronic noise in the pixel ramps. 

Currently no reference pixel correction is made for subarray data, even when the subarray includes the left column of reference pixels. Since subarrays are typically used for short integrations, the effects of electronic drifts are small and there is less need to correct for them.

\subsection{Jump Detection}
\label{jumps}
An extensive analysis of the effects of cosmic ray hits on the \textit{Spitzer} IRAC Si:As IBC detectors can be found in \citet{hagan2021}. We assume that this analysis applies to the MIRI detectors also, since they are very similar. 

The signal produced by a hit arises through ionization of the detector material in the IR-active layer. This layer is sufficiently thin that the statistics of this process produce a broad spectrum of signals. There is a tail toward high energies following the Landau Distribution and a broad distribution toward small signals. In addition, signals from a hit pixel are impressed on neighboring ones through a combination of electrical cross-talk through interpixel capacitance (IPC) and migration of thermal and super-thermal free electrons \citep{hagan2021}. 

\begin{figure*}
\begin{center}
\begin{tabular}{c}
\includegraphics[width=0.95\textwidth]{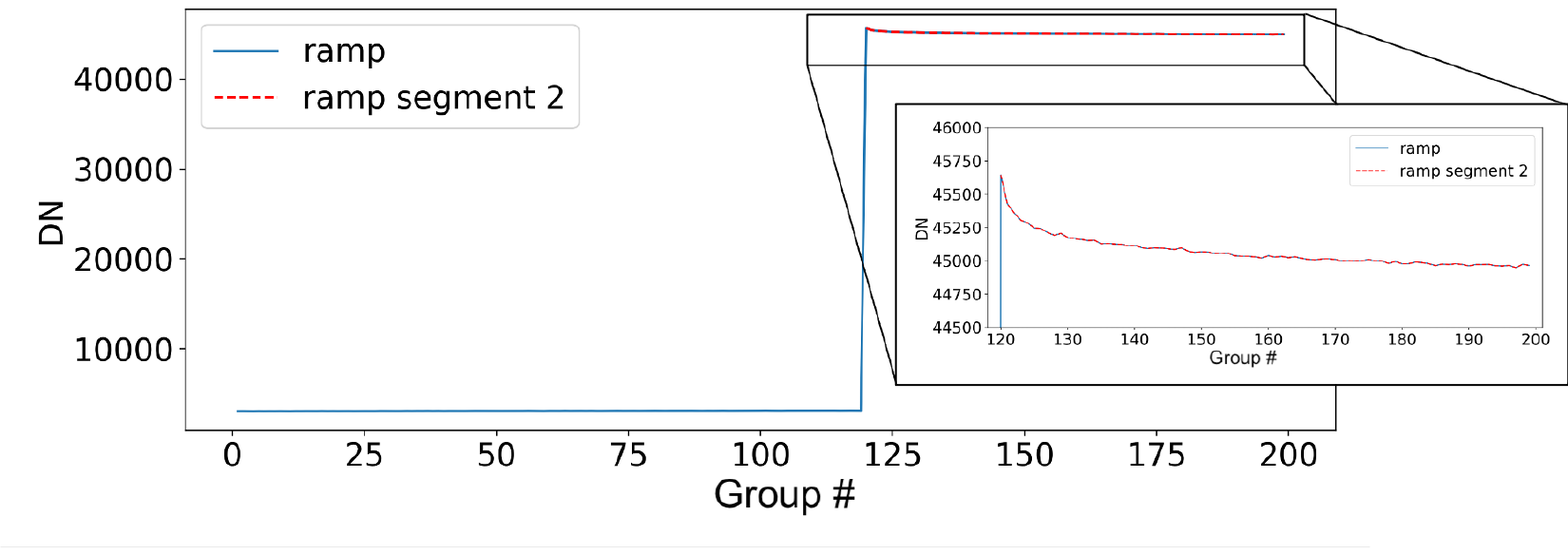}
\end{tabular}
\end{center}
\caption 
{ \label{fig:cosmic_charge_decay} {Dark measurement for a single pixel with a large cosmic ray hit. The integration ramp values after a cosmic ray hit show a negative trend (see inset).}}
\end{figure*}

Analysis of the in-flight MIRI data has shown a variety of ramp jump signatures due to cosmic ray hits. Ramp jumps that only affect one or a few pixels are seen to be impressed on neighboring pixels even below the detection threshold in a single ramp. We have also seen a class of effects for large hits that impacts a large number of pixels, termed "showers" and that have a generally elliptical shape.  In addition, the large cosmic ray hits show a short-lived transient that impacts the signal for tens of seconds after the ramp jump. Figure \ref{fig:cosmic_charge_decay} shows how a large cosmic ray hit has an impact on the ramp after the hit.  After running the data through the \textsc{calwebb\_detector1} pipeline, the ramp after the CR hit shows a negative count rate value.

\subsubsection{Cosmic Ray detection in the JWST Pipeline for MIRI Data}

\begin{figure}
\begin{center}
\begin{tabular}{c}
\includegraphics[width=0.45\textwidth]{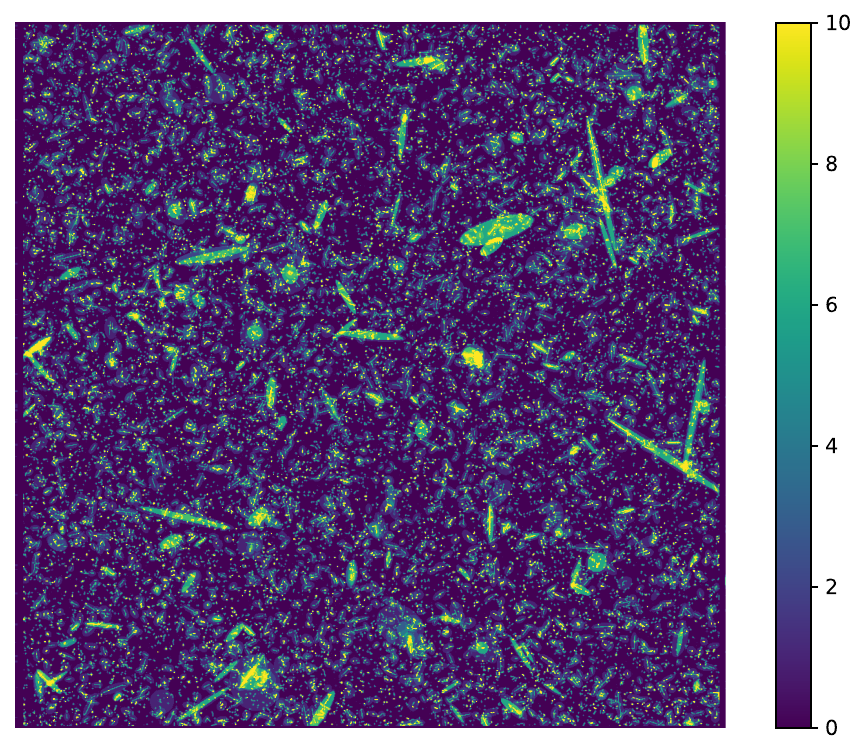}
\end{tabular}
\end{center}
\caption 
{ \label{fig:cosmic_ray_showers} { An image of the number of groups flagged as affected by cosmic rays  in a 250 group MIRI dark exposure.}}
\end{figure}

The initial flagging of ramp jumps is done using the 2pt difference algorithm \citep{anderson2011}. This flags the group before the ramp jump, telling the ramp fitting step to perform a linear fit before and after this group and to average the resulting slopes. The initial ramp jump flagging is expanded to the same group in neighboring pixels to account for pixel cross-talk and to a few groups after the jumps to account for the short-lived temporal transient. In addition, to account for showers, a preliminary spatial detection algorithm finds co-spatial regions with initially detected jumps, fits the region with an ellipse, expands the ellipse by a set fraction, and flags all the pixels in the larger ellipse. Updates and improvements to this preliminary algorithm are currently being studied. Figure \ref{fig:cosmic_ray_showers} shows the number of groups flagged as affected by cosmic rays in a 250 group dark exposure.

\section{Results}
\begin{figure*}
\begin{center}
\begin{tabular}{c}
\includegraphics[width=0.95\textwidth]{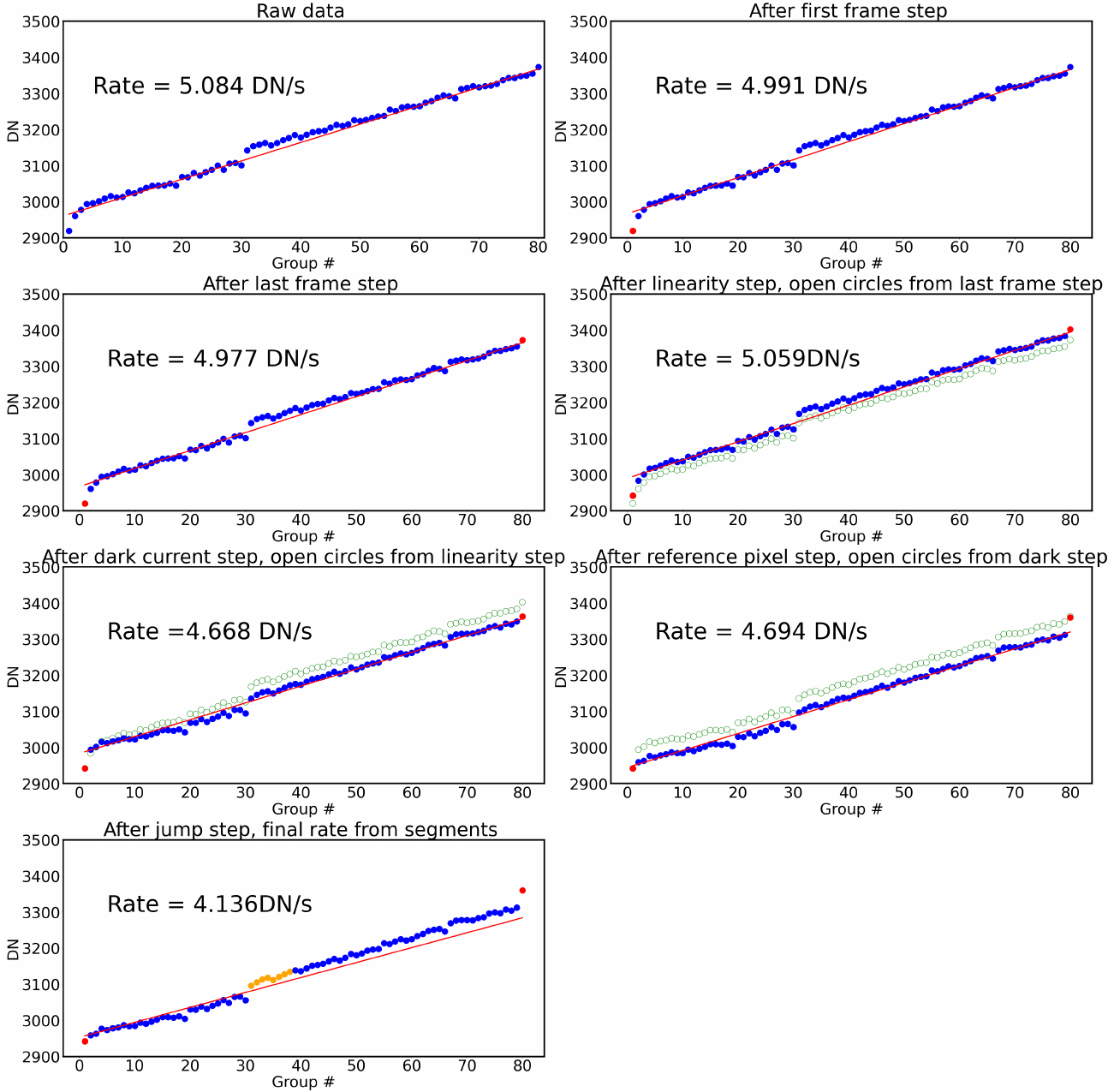}
\end{tabular}
\end{center}
\caption 
{ \label{fig:mrs_example} { An MRS pixel ramp with 80 groups in an integration. Each panel shows the results of specific steps of the {\sc calwebb\_detector1} pipeline. The rate determined by the ramp fitting step using the data plotted in the blue circles is shown in each plot. Red points represent groups that are skipped at the beginning or end of an integration, orange points represents groups skipped by the jump detection step. Green circles are the results of the previous step shown as a reference. The final rate  after the jump step is determined from two weighted segments between the groups affected by the cosmic ray.}}
\end{figure*}

\begin{figure*}
\begin{center}
\begin{tabular}{c}
\includegraphics[width=0.95\textwidth]{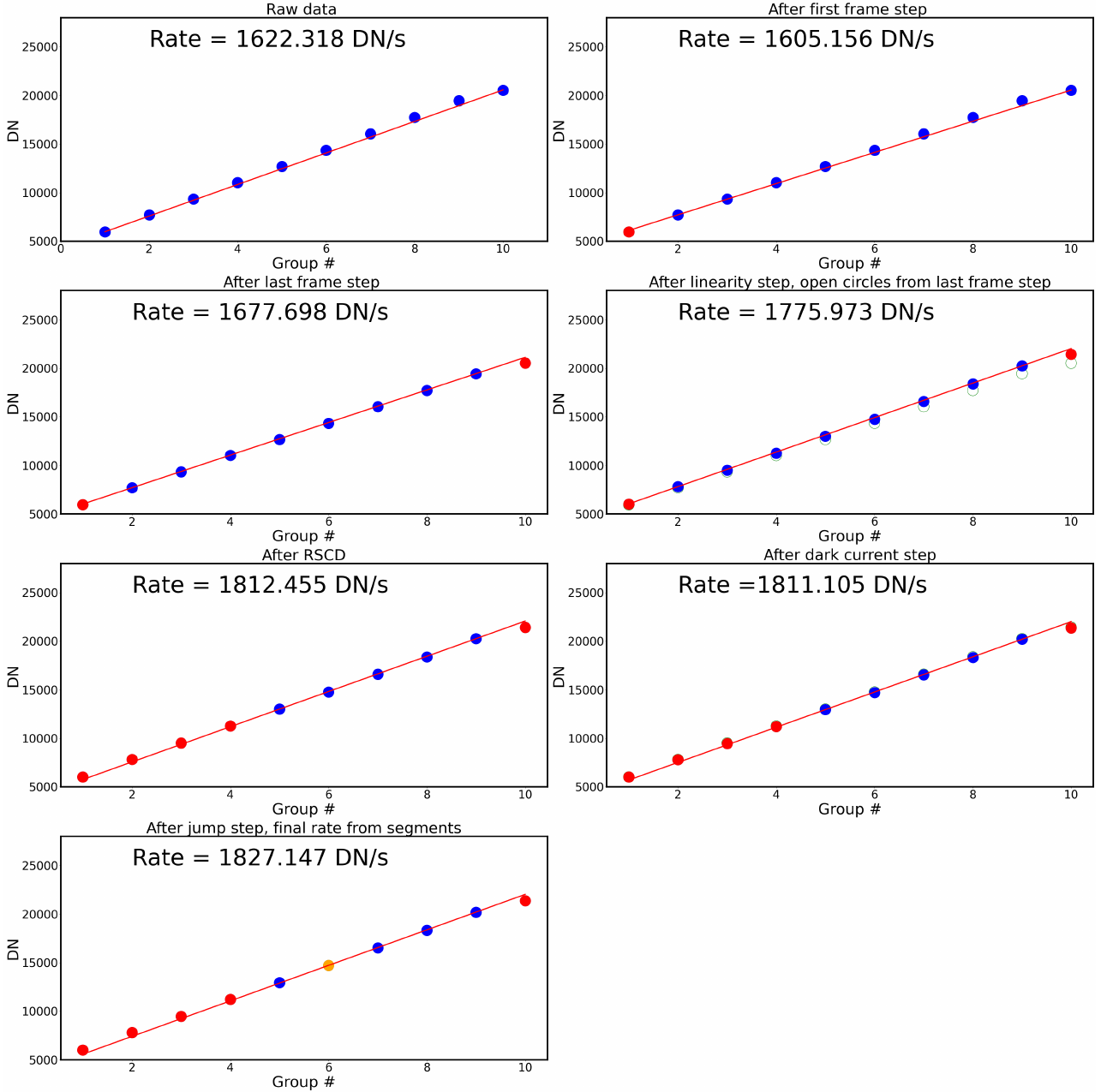}
\end{tabular}
\end{center}
\caption 
{ \label{fig:subarray_example} {An Imager Sub256  10 group pixel ramp for the second integration. The second integration is shown to illustrate the rejection of groups in  the RSCD step. Each panel shows the results of specific steps of the JWST pipeline. The rate determined by the ramp fitting step using the data plotted in the blue circles and the resulting rate is shown in each plot. Red points represent groups that are skipped at the beginning or end of an integration or from the RSCCD step, orange points represents groups skipped by the jump detection step. The last plot shows group 6 is flagged as a cosmic ray. The final rate in the case will be determined by groups 7, 8, and 9.}}
\end{figure*}

In this section we show the results on a pixel ramp of each step discussed in the preceding section. The count rate (slope) given on each plot is the result of running the {\sc calwebb\_detector1} pipeline up to that step and using the ramp fitting step to derive a slope value. The terminology used by the JWST pipeline for the end product of the {\sc calwebb\_detector1}  pipeline is rate (instead of slope), hence for these plots we use the term rate.  We have shown two types of results: a faint long MRS exposure in  Figure \ref{fig:mrs_example} and a short bright subarray Imager integration in Figure \ref{fig:subarray_example}. In both cases the first and last groups are not used in the linear fitting of the ramp. Blue points are valid group values for the ramp, red points are rejected groups, and orange values are groups flagged as affected by a cosmic ray. In Figure \ref{fig:mrs_example} we have plotted the results of the previous step as open green circles to highlight the differences from the results of the  current step shown in the panel.  An obvious jump resulting from a cosmic ray is flagged as seen in Figure \ref{fig:mrs_example}. The amplitude of the cosmic ray was high enough to cause the flagging of a total of  eight groups. When a jump is detected, the ramp is split into a segment before and after the flagged groups. The final rate determined for the ramp is a weighted mean of the linear fit to each segment. Figure \ref{fig:subarray_example} is  a ten group SUB256 exposure for the second integration. We have shown the results of each step for the second integration to highlight the rejection of groups in the RSCD step.  Only the panel showing the results of the linearity correction also shows the results of the previous step (last frame correction), because most the steps are only rejecting groups or the previous step has a small effect because the data are for a bright source. The last plot shows the points used to derive the final rate for the integration. A jump has been detected in group 6, which does not seem obvious from the ramp. This group is flagged because a neighboring pixel had a cosmic ray hit on group 6. If a pixel's ramp is affected by a cosmic ray, the pipeline will (by default) also flag the four perpendicular neighbors of
this central pixel as also having a jump. This is needed because of the inter-pixel capacitance  causing a small jump in the neighbors. The small jump might be below the rejection threshold but will affect the slope determination of the pixel. In the default processing only  groups 7, 8 and 9 will be used in the linear fit to derive the final integration rate for this pixel, since the segment before the jump has only one group. 

\section{Data Artifacts Not Removed in the Pipeline} \label{NoCorrection}

\subsection{Persistence}
The MIRI detector arrays exhibit lingering images after a bright source has been removed, a common artifact found in cryogenic infrared detectors, that is referred to as "persistence" \footnote{Persistence as discussed in this paper is also commonly referred to as latents/latency but we, and the JWST documentation in general, avoid the word “latency” because it is also associated with properties that cannot be attributed to the persistence effect we describe.} in the JWST documentation. Commissioning analysis has shown that the persistence decay times in flight are typically a few minutes up to half an hour, where the amplitude of the persistence depends both on the luminosity of the source of origin and the time observing that source. Ground testing has shown the persistence to be well modelled by multiple time constants, suggesting that there are a number of mechanisms that contribute to the effect. The first potential cause, and one common to most solid state detectors, is traps that can capture free charge carriers and release them slowly into the detector volume. The all-silicon construction of the Si:As IBC detectors leaves only two surfaces where traps are likely: (1) at the frontside silicon-to-metal contact interfaces; and (2) at the buried contact. A graded electrical interface is provided at the frontside contacts by ion implantation and the buried contact is also formed in this way \citep{love2004}. The residual damage from this process is partially removed by high-temperature annealing , but traps will remain in both places (see \cite{Dicken2023} for details of characterizing the  MIRI Persistance in Commissioning). A second possibility for persistent images is slow dielectric relaxation in the RC circuit across the detectors (M.``Dutch'' Stapelbroeck, private communication).

The pattern of persistence can be complex: for IRAC, there were differences between the two Si:As IBC arrays, and for one of the arrays in addition to classical persistent images that decayed in minutes, there were long-term ones that could last for weeks (see IRAC Instrument Handbook \footnote{https://irsa.ipac.caltech.edu/data/SPITZER/docs/irac/iracinstrumenthandbook/}). However, the MIRI detector persistence characteristics from commissioning results appear to be very favorable compared to previous missions using the same detector technology. Not only is the persistence relatively low compared to previous missions, but its decay function for a given luminosity is also very homogeneous across the detectors.

Because the persistence measured in flight is seen to be a small fraction of the original illuminating flux ($<$0.01\%), it is not thought to be of concern for most MIRI users. Persistence affects all of MIRI's detectors but is more likely in MIRI imaging than the MRS or LRS due to its on average higher throughput e.g. when a bright target is slewed across the field of view. However, persistence will be seen in all MIRI detectors from high power cosmic rays and showers, although pipeline corrections are being improved to flag several groups after such cosmic rays. See \cite{Dicken2023} for examples and further discussion on persistence in MIRI imaging. 

\subsection{Cruciform}\label{cruciform}

There is a feature caused by internal scattering within the MIRI detectors that manifests as a cruciform shape in imaging and spectroscopy at wavelengths $\leq$ 10 $\mu$m.  This behavior was modeled in detail in \citet{gaspar2021}. This effect is seen in the rate images  resulting in a classic cruciform-shaped diffraction pattern that follows the row and column directions of the detector. The cruciform is an optical effect and is not removed by {\sc calwebb\_detector1}. In the case of the MIRI imaging, this effect is left to the user to model and remove if doing so is necessary for their particular science case.  For more detail on the Imager and the cruciform effect see \cite{Dicken2023}.  This effect is more significant in the case of the MIRI MRS for which the interleaving of the IFU slices means that the cruciform artifact contributes `bands' of emission that are non-local on the sky in the reconstructed data cubes \cite{Argyriou2023}. The MIRI MRS pipeline therefore models and removes the cruciform artifact in the {\sc calwebb\_spec2} pipeline by convolving the observed data with a series of templates constructed from observations of bright point sources, as discussed in more detail by \cite{Patapis2023}.

\subsection{Interpixel Capacitance}

As with most solid-state detector arrays, there is a low level of interpixel capacitance (IPC) that couples signal accumulated on one pixel to its neighbors. The resulting cross-talk for the MIRI arrays is $\sim$ 3\% for the pixels sharing an edge with the one receiving the signal, and much less, but not zero, for those at the corners. Two approaches can be used for the IPC crosstalk: (1) correct for it in the pipeline; or (2) treat it as an imaging issue and correct for the slight broadening of image in post-pipeline processing, if desired. The strongest argument for the first approach is that, in principle, as a purely electrical effect, it might be possible to use an IPC correction to remove the effects of cosmic ray hits on pixels surrounding the one hit. However, analyses of the muon hits in laboratory data showed that the signals induced in surrounding pixels were not sufficiently repeatable or predictable for this approach to be effective. A highly accurate removal might also be difficult, since small layout variations in the detector array might result in variations in the IPC across its face. We therefore leave dealing with IPC as a post-pipeline activity.

\subsection{Column and Row artifacts}
 
\begin{figure*}
\begin{center}
\begin{tabular}{c}
\includegraphics[height=8.0cm]{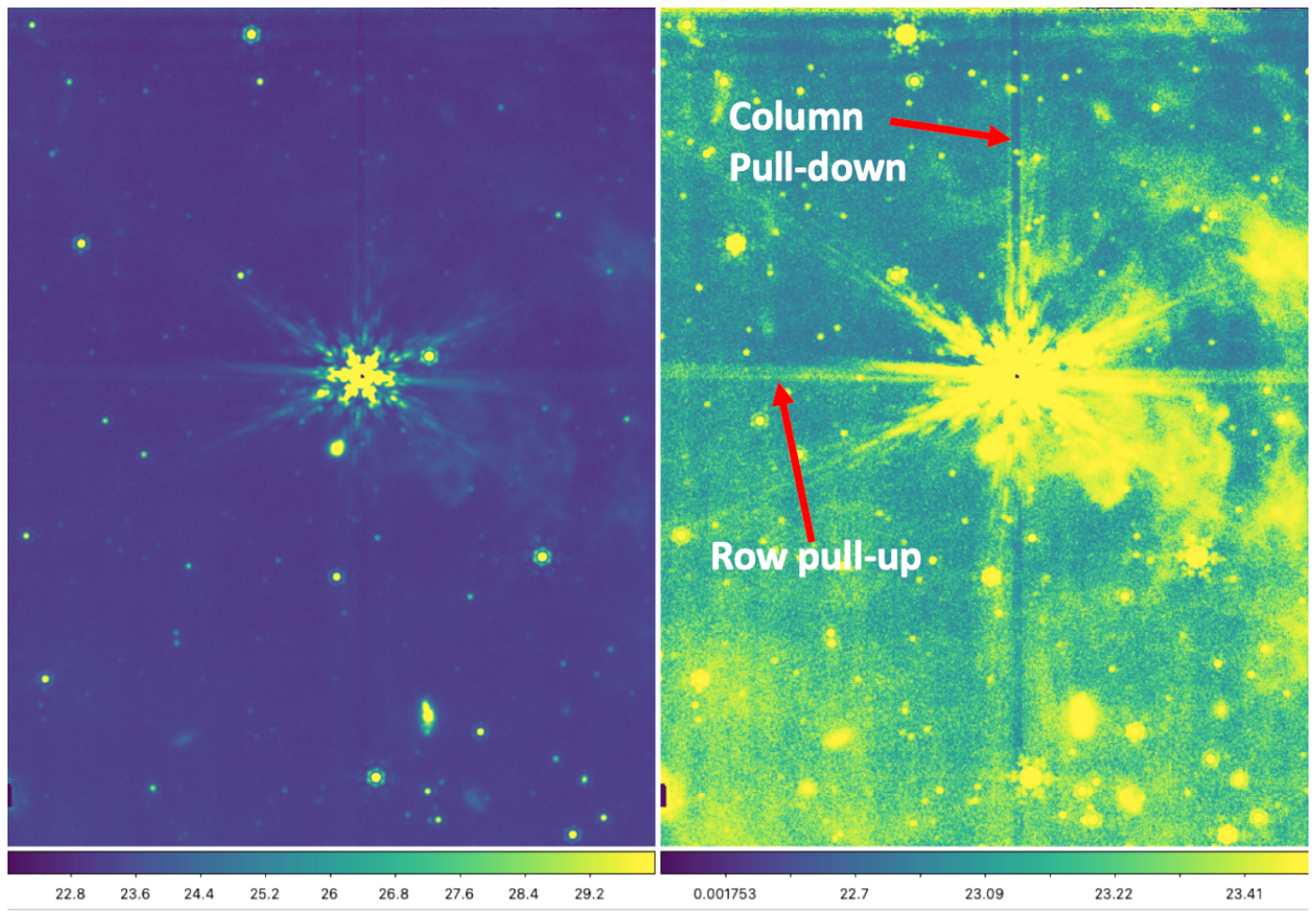}
\end{tabular}
\end{center}
\caption
{ \label{fig:row_column} {A MIRI image at 12.8 $\mu$m of a bright planetary nebula source in Program 1090. The left image shows the source; the right image is identical but with the background flux stretched to reveal the column (pull-down) and row (pull-up) artifacts.}}
\end{figure*}

 Similar to the Spitzer IRAC (IRAC Instrument Handbook Section 7.2.4), the MIRI images show a row-column effect in which a bright source on the detector can pull up or pull down the flux value in the entire rows or columns that the source is observed in (See Figure \ref{fig:row_column}).  We will give a top level summary of this effect, for details on this effect see \citet{dicken2022}. Although the row and column effects are more apparent in imaging they also affect the MRS. The MRS disperses two channels on one detector and a bright emission line can cause a row effect artifact that pulls up or down the pixels in the neighboring channel \citet{Argyriou2023}.

From ground testing we know that row and column artifacts are present at all illumination levels, but they are more noticeable when the observed source is at high contrast compared to the background.  The effect is seen in rate images (stage 2 and above in the pipeline) but the underlying cause of the artifact is a change in the shape of the ramp in level 1 data. Although still under investigation, we know this change in the ramp shape is related to how quickly a high contrasting source in any row or column is saturating. Because of this, the length of the ramps is an important factor in determining if the flux change is positive (pulled up) or negative (pulled down). Also we know that the column artifacts are only apparent in columns that include bright contrasting sources, whereas the row effect is also seen in rows above those that include the sources - in the read direction of the detector. Pipeline solutions for the row and column artifacts are currently under development either as a filtering/cleaning tool in stage 2 as existed for Spitzer or as a correction in stage 1 data.  
 
\subsection{First Exposure Effect}

The first integration of the first exposure of a dithered data set can have a slope that is offset from the other integrations. The MIRI team has termed this difference the \textit{the first exposure effect}.  As mentioned earlier,  the MIRI detectors are clocked out at a constant rate whether observing or not. The pixels are continuously addressed at time intervals of 10~$\mu$s. In between dithered exposures there is a short time interval to set up the exposure, but between visits there is a much longer time interval where the detector is in idle reset. This difference in idle time results in a difference in the rate in first integration in the first exposure from the other first integration rates from the dithered data.  Currently this is an open issue in the JWST pipeline. One possible correction could use the regions on the detector not directly illuminated (i.e. behind the mask in the Imager or between the channels on the MRS detectors) to normalize the count rates.

\subsection{Brighter-Fatter Effect}\label{bfe}

\begin{figure*}
\includegraphics[width=\textwidth]{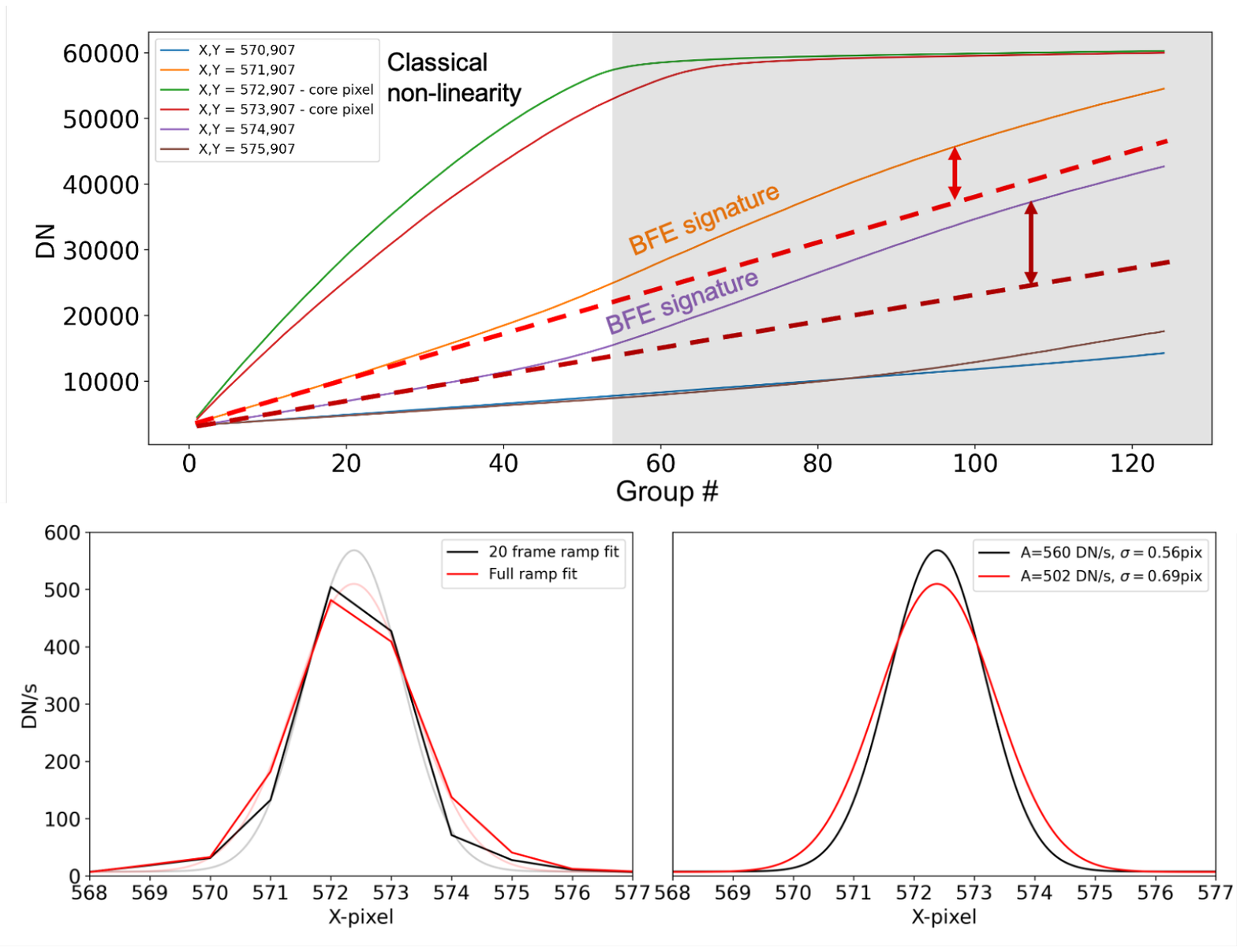}
\caption 
{ \label{fig:bfe_miri} {Top: Plot illustrating how the brighter-fatter effect (BFE) traditionally considered in classical CCDs and H2RG detectors manifests in MIRI detector pixel integration ramps. Bright pixels (X=572 and X=573) experience a "classical" non-linearity. Pixels measuring fainter signal and neighboring the bright pixels show ramps that deviate from the classical model. Overplotted dashed lines accentuate the BFE signature. Bottom: MIRI Imager PSF based on pixel ramp slope fits (left: 20 groups versus full ramp). 1D Gaussian profiles are fitted to the two PSFs; right. The MIRI BFE results in a 23\% broader PSF in the case where the full ramp is used for the slope fitting.}}
\end{figure*}

MIRI integration ramps show a prominent type of cross-talk, mostly visible at medium and high fractions of the pixel full-well capacity. The effect is reminiscent of the brighter-fatter effect (BFE) in traditional CCDs and H2RG detectors \citep{Coulton2018,Plazas2018}. The impact of this type of cross-talk on MIRI ramps is shown in Fig.~\ref{fig:bfe_miri} where a horizontal cut is taken through a bright point source observed with the MIRI Imager, such as the one shown in Fig.~\ref{fig:row_column}. The signal in the ramps is plotted in DN against group number. We can see that for the two pixels that sample the core of the point-spread function (PSF), represented by the green and the red solid curves, the shape of the ramps adheres to the traditional drop in pixel response linked to the debiasing of the MIRI detectors ("classical" non-linearity). Looking at one pixel to the left of the PSF core (pixel X=571), instead of a drop in response and a decrease in slope, the ramp shows a steady increase in slope. We overplot straight dashed lines to accentuate the BFE signature; in the case of classical non-linearity, the signal in the ramps should never go above the straight line. Instead, there is a gradual increase in the slope of the measured ramp starting from low DN values, that reaches a maximum gradient when the PSF core pixel (X=572) reaches full-well. Afterwards the ramp appears to behave in a "classical" non-linear manner. A similar relation is shared between the PSF core pixel (X=573) and the PSF wing pixel (X=574). Naturally a PSF is two-dimensional, hence the effect is in reality two-dimensional as well.

Since the JWST pipeline flux estimation is based on applying a non-linearity correction to the ramps and determining the slopes of said ramps, the non-linearity correction amplifies the BFE signature by "bending" the impacted ramps further upwards. If the effect is not corrected or mitigated, the final estimated slope ends up being systematically larger. The result is an artificially broader PSF FWHM as a function of (1) the point source brightness and (2) the length of the integration ramps (i.e., the fraction of the full well reached by the pixels). In the bottom two panels of Fig.~\ref{fig:bfe_miri} we show the effect on the PSF by comparing the ramp fitting results of different numbers of groups, after a non-linearity correction has been applied. In the first case only 20 groups are used to determine the ramp slope, and in the second case all the groups up to saturation are used ("full ramp"). The resulting fitted Gaussian profiles are compared in the bottom right panel. BFE reduces the peak amplitude of the PSF, and increases the FWHM (broadens the PSF) by $\sim$20\%.

To mitigate the impact of the brighter-fatter effect, the JWST pipeline flags the pixels that reach a DN stage above the saturation limit and then flags the groups of all surrounding pixels past the group where the bright pixels saturate. These flagged groups are not used for the slope determination of the surrounding pixels. This is illustrated in the top panel of Fig.~\ref{fig:bfe_miri} by the grey shaded area. The pixel X=572 saturates first at group number 56, hence for pixels X=571 and X=573 only the part of the ramp up to group number 56 is used to fit the slope. Importantly, even with this mitigation strategy, the BFE still broadens the PSF by $\sim$10\%. This must be taken into account when performing optimal PSF-weighted photometry.

The BFE impacts MIRI imaging and spectroscopy alike. Due to its complexity, no analytical model has been developed to this date for the MIRI BFE.  In the case of spectroscopic measurements of very bright emission lines, the line spread function (LSF) will be broadened in a similar fashion to the PSF shown in Fig.~\ref{fig:bfe_miri}. An examination of the raw ramps is therefore recommended for bright sources observed with MIRI.

\section{Summary}
This paper has described the MIRI specific {\sc calwebb\_detector1} processes in the JWST pipeline. The data reduction algorithms and derivation of the reference files have been guided by extensive ground based testing using engineering and flight arrays and now flight experience. 
Before the launch of JWST we did not appreciate the full impact of having  time-resolved cosmic ray showers. We have no evidence that the cosmic ray effects on the MIRI detectors are different from those impacting similar space-based detectors (i.e., the IRAC Si:As IBC detectors).  The main difference in terms of dealing with cosmic rays is the type of  data downlinked from the telescopes. The MIRI detectors, when run in FAST mode, downlink all the non-destructive reads (groups) for each pixel in an integration; while the IRAC detectors used an on-board differencing of multiple reads and only downloaded one integer value per pixel per exposure. The experience gained from having flight data consisting of the several non-destructive reads in an integration has allowed  new algorithms to be  developed to try and mitigate the effects of the showers. The design and operation of the MIRI instrument has been summarized to provide sufficient background for the understanding of the data reduction algorithms. We have provided specific details on how the readout process and design of the instrument has influenced the different components that need to be included in algorithms. Initial testing using flight data has validated these data reduction algorithms, but studies continue to improve upon each step in the pipeline.

\begin{acknowledgments}
Ioannis Argyriou and Danny Gasman thank the European Space Agency (ESA) and the Belgian Federal Science Policy Office (BELSPO) for their support in the framework of the PRODEX Programme. Pierre Guillard would like to thank the Sorbonne University, the Institut Universitaire de France, the Centre National d'Etudes Spatiales (CNES), the "Programme National de Cosmologie and Galaxies" (PNCG) and the "Physique Chimie du Milieu Interstellaire" (PCMI) programs of CNRS/INSU, with INC/INP co-funded by CEA and CNES, for their financial support. The work of J. Morrison, G. Rieke, S. Alberts, A. G\'asp\'ar, and I. Shivaei was supported in part by NASA grants NNX13AD82G and 1255094.  Javier \'Alvarez-M\'arquez acknowledges support by grant PIB2021-127718NB-100 from the Spanish Ministry of Science and Innovation/State Agency of Research MCIN/AEI/10.13039/501100011033 and by “ERDF A way of making Europe”.  This work was co-authored by an employee of Caltech/IPAC (Seppo Laine) under Contract No. 80GSFC21R0032 with the National Aeronautics and Space Administration."
\end{acknowledgments}


\bibliography{bib_list}   
\bibliographystyle{aasjournal}

\end{document}